\documentclass[a4paper,11pt]{article}
\pdfoutput=1 

\usepackage{jcappub} 

\usepackage[T1]{fontenc} 

\title{Implications of multi-axion dark matter on structure formation}

\author{Chong-Bin Chen$^\dag$,}
\author{Jiro Soda$^{\dag,\sharp}$}

\affiliation{$^\dag$ Department of Physics, Kobe University, Kobe 657-8501, Japan\\
 $^\sharp$ International Center for Quantum-field Measurement Systems for Studies of the Universe
 and  Particles (QUP), KEK, Tsukuba 305-0801, Japan}

\emailAdd{chongbin@stu.kobe-u.ac.jp}
\emailAdd{jiro@phys.sci.kobe-u.ac.jp}

\abstract{Axions are candidates for dark matter in the universe.We develop an accurate Boltzmann code to calculate the linear growth of the plasma.
As an interesting example, we investigate a mixed dark matter model consisting of cold dark matter (CDM) and two-axion dark matter.  We analyze the growth of the structure numerically and analytically. We find that an effective single axion with an effective mass and an effective abundance is useful to characterize the two-axion cosmology. Moreover, we generalize the effective single axion description to multi-axion dark matter cosmology. 
We also compare the results with those of warm dark matter (WDM) model. 
Moreover,  we calculate halo mass functions for the mixed model by using the Press-Schechter model and linear perturbations and then determine the
mass function as a function of masses and axion abundance.
}

\begin{document}
\maketitle
\flushbottom

\section{Introduction}
\label{sec:Intro}
One of the most successful model of cosmology is the so-called $\Lambda$CDM model, which includes cold dark matter(CDM) and dark energy together with matter in the Standard Model of particle physics.  In fact, $\Lambda$CDM model has exhibited excellent agreements with current observations on large scales  \cite{Planck:2018jri,BICEP2:2015xme,BOSS:2013rlg}. 
Among them, CDM  is responsible for the formation of structure. However,  CDM has gradually revealed problems in understanding small-scale structures (see the review article \cite{Bullock:2017xww}). These problems are particularly 
serious in dwarf galaxies. The N-body simulation revels a universal density profile, the NFW profile of the dark matter haloes \cite{Navarro:1996gj}. But the actual density profile  of low-mass galaxies is cuspless than the profile predicted. The density approaches almost constant near the central regions of galaxies. Also, the numbers of observed satellites and haloes with mass $10^{10}M_{\odot}$ is much smaller than the theoretical perdition. One possible reason of this discrepancy is the lack of feedback of baryons to the dark matter halos \cite{Governato:2012fa}. However, modelling all baryonic processes is still challenging and baryonic effects may be insignificant for ultra-faint dwarfs \cite{Bullock:2017xww, Tollet:2016,Fitts:2016usl,Hopkins:2017ycn}. Another solution is that the dark matter behaves quite differently from CDM on small scales, i.e., their collapse should be suppressed on small scales, hence 
the density profile is flattened and the number of small structure formation is decreased. 

The mitigation of small scale issues can be responsible by warm dark matter(WDM), which washes out small scale fluctuations because of the free-streaming. However,  WDM has the $Catch\ 22$ problem, i.e.,  WDM mass constrained by large scale structure can not create desired size of core in the dwarf galaxy \cite{Maccio:2012qf}. The small scale issues can be also resolved by the so-called Fuzzy dark matter(FDM) \cite{Hu:2000ke} or wave dark matter($\psi$DM) \cite{Schive:2014dra}. Indeed, ultralight scalar bosons with a galactic size($\sim$kpc) de Broglie wavelength can mimic CDM on large scales while they behave as wave-like matter on small scales. The mass of ultralight dark matter (ULDM) can be constrained by the large scale CMB anisotropy to $m\gtrsim 10^{-24}$eV if most of dark matter is composed of ULDM \cite{Hlozek:2014lca,Marsh:2015xka,Hlozek:2016lzm}. For  small scales, the simulations of a solitonic core of ULDM with the non-relativistic Schr\"{o}dinger-Poisson equations  suggest that ULDM should have $m\sim10^{-22}$eV to solve the cusp-core crisis \cite{Hu:2000ke,Schive:2014dra,DeMartino:2018zkx,Schive:2014hza}. On the other hand, a more low-cost method is solving the hydrodynamic Madelung equations of ULDM \cite{Veltmaat:2016rxo,Mocz:2015sda,Nori:2018hud,Zhang:2016uiy}. From hydrodynamics point of view, the wave nature of ultralight dark matter(ULDM) is described by a cut-off scales and the ``quantum'' pressure. The results of hydrodynamic simulation can be compared with the Lyman-$\alpha$ forest data, which is a useful method for probing the linear power spectrum on small scales ($0.5\text{Mpc}/h\sim 10\text{Mpc}/h$) \cite{Irsic:2017yje,Kobayashi:2017jcf,Veltmaat:2016rxo,Armengaud:2017nkf}. The hydrodynamic simulation together with Lyman-$\alpha$ forest data has ruled out the ULDM mass $m\lesssim 2\times10^{-20}$eV at $95\%$ confidence level \cite{Rogers:2020ltq}. Recently, the constraints on subhalo mass function from strong gravitational lensing of quasar and fluctuations in stellar streams also ruled out the mass range $m\lesssim2.1\times 10^{-21}$ assuming all dark matter is made up of ULDM \cite{Schutz:2020jox}. The mass ranges constrained from the power spectrum seem too heavy to relieve the cusp-core crisis. This inconsistency of ULDM models needs to be taken seriously.

The so-called extreme axion model was proposed to mitigate this problem \cite{Leong:2018opi}. Axion is one kind of candidates of the ULDM and can be produced by spontaneous breaking of Peccei–Quinn(PQ) symmetry, which is initially proposed to solve the strong CP problem in QCD. And then such axion particles are considered as DM candidates \cite{Preskill:1982cy,Abbott:1982af,Dine:1982ah}. In string theory, such kinds of axion-like particles with ultralight mass, which is called ultralight axion(ULA) in this paper, can be produced in low-energy effective theory. In the extreme axion model, the ULA should stay very close to the hilltop of the periodical potential at the beginning.  However, the initial location of ULA is randomly distributed if the symmetry breaks during inflation.  The mass of ULA is determined by decay constant and instanton scale
\begin{equation}\label{mass}
    m^2=\frac{\Lambda_{a}^4}{f_{a}^2}.
\end{equation}
The scale of instanton effect is given by $\Lambda_{a}^4=\mu^4e^{-S}$, where $\mu^4$ is the UV energy scale related to SUSY breaking and Planck scale and $S$ is the area of the compact cycles. The sting compactification results in hundreds of closed cycles,  hence the ULA mass is logarithmically distributed  over a very wide mass range. Therefore it is natural to consider the universe in the presence of multiple ULAs with different mass and different initial conditions \cite{Arvanitaki:2009fg}. Other than theoretical considerations, recently some signatures of multiple ULAs in our universe were found. In \cite{Emami:2018rxq} they claimed  that there may be evidence for light bosons with more heavier mass $m>10^{-18}$eV to 
be responsible for the solitonic core of 47-Tuc globular cluster. Also, in the Milky Way, the analysis of ultra-faint dwarf galaxies suggests a double-soliton structure of ULAs on different scales corresponding to $10^{-22}$eV and $10^{-20}$eV, respectively \cite{Luu:2018afg}. Moreover, recently the analysis of cores of a range of galaxies reveals the relation of core density and core radius \cite{Deng:2018jjz}. However, most of theoretical models of a single ULA fail to predict this relation. It has been pointed out that the galactic condensates of multiple ULAs including self-interaction can relax this issue \cite{Eby:2020eas,Guo:2020tla}. The simulation of the galactic core of multi-ULA is also being studied recently \cite{Huang:2022ffc,Gosenca:2023yjc,Glennon:2023jsp}.

On the other hand, the constraint by Lyman-$\alpha$ forest on linear power spectrum, which is in tension with density profile simulation for a single ULA, should be addressed. This is important for simulations of the ULAs in the non-linear regime.  Constraints on multiple ULAs from large scale linear perturbations have been investigated in \cite{Hsu:2020ikn}. The multiple ULAs may also solve the problem of the inconsistency from Lyman-$\alpha$ constraints in extreme axion models \cite{Tellez-Tovar:2021mge}. However, the Lyman-$\alpha$ forest constraints are very sensitive to the shape of the transfer function of linear power spectrum because the halo mass scales as radius cubed. Therefore, it is necessary to systematically study how multiple ULAs deform the shape of linear power spectrum. This is what we are going to do in this paper. Since the structure may be formed by mixed dark matter \cite{Marsh:2013ywa,Vogt:2022bwy}, we consider the universe with  ULAs and CDM, where ULAs are only a small fraction of the total dark matter. There are two important parameters, $m_{r}\equiv m_{1}/m_{2}$ and $\omega_{r}\equiv \Omega_{a1}/\Omega_{a2}$ in the two-ULA case. We analytically compute and see how transfer functions and halo mass functions depend on these parameters. Then we present numerical results for different fractions of ULAs. It turns out that effective single axion description is useful
for two ULAs. Remarkably, we find a general formula for the effective mass in the multi-ULA cases. We also compare the results with those of WDM. Moreover, we will calculate halo mass functions for two-ULA dark matter.

The organization of the paper is as follows.  In section \ref{sec:ULAphysics}, we review some relevant equations and quantities of ULAs on cosmological scales. In section \ref{sec:Boltzmanncode}, we introduce our Boltzmann code for accurate calculations of power spectrum in this paper. In section \ref{sec:Suppression}, we study how ULAs suppress the linear matter power spectrum on small scales and also compare with  WDM.
We show the effective single axion picture is useful for analyzing multi-axion dark matter.  
In section \ref{sec:haloes}, we discuss the dark matter haloes model in the presence of ULAs. We analytically and numerically compute the variance and mass functions. The final section is devoted to the discussion.

\section{Ultralight axion cosmology}
\label{sec:ULAphysics}
The low-energy effective Lagrangian of ULA is given by
\begin{equation}\label{Lagrangian}
\mathcal{L}=-\frac{1}{2}f_{a}^2 \left(\partial \Theta\right)^2-\Lambda_{a}^4U(\Theta),
\end{equation}
where $f_a$ is the PQ symmetry breaking scale of a global $U(1)$ symmetry of ULA and $\Lambda_{a}^4$ is the energy scale of instanton effect. This effect breaks the shift symmetry of ULA to a discrete one 
and leads to a periodic potential $U(\Theta)$
with a minimum at $\Theta=0$. After canonically normalizing the field $\phi=f_{a}\Theta$, one can expand the potential near minimum with mass (\ref{mass}). From Lagrangian (\ref{Lagrangian}), the background equation of motion of ULA $\phi$ in a Friedmann-Robertson-Walker universe is obtained as
\begin{equation}\label{axioneom}
    \phi''+2\mathcal{H}^2\phi'+a^2m^2\phi=0,
\end{equation}
where $\mathcal{H}(a)=aH(a)$ and $H(a)$ is the Hubble parameter. Here a prime denotes the derivative with respect to the conformal time $\tau$. At the early stage, the Hubble friction is much larger than the ULA mass, so the ULA field $\phi$ behaves as DE. While at the late stage, the Hubble expansion rate becomes smaller than the mass, so ULA starts to oscillate with a frequency $\sim m$. In this stage, the energy density of ULA evolves like CDM. This transition occurs roughly at 
\begin{equation}\label{Osctime}
    H(\tau_{m})=m.
\end{equation}
The scalar field equation of motion can be rewritten as a conservation equation of generalized DM as \cite{Hu:1998kj,Hu:2003hjx}
\begin{equation}\label{axioneffeom}
    \rho_{a}'=-3\mathcal{H}\left(1+w_{a}\right)\rho_{a},
\end{equation}
where $w_{a}\equiv P_{a}/\rho_{a}$ is the equation of state of ULA and
\begin{align}
    &\rho_{a}\equiv\frac{1}{2a^2}\left(\phi'\right)^2+\frac{m^2}{2}\phi^2,\ \ \ \ \
    P_{a}\equiv\frac{1}{2a^2}\left(\phi'\right)^2-\frac{m^2}{2}\phi^2.
\end{align}
The equation of state also oscillates after $\tau_{m}$. At early stage $\phi$ is a constant so the energy density $\rho_{a}$ redshifts as DE. After transition time $\tau_{m}$ the ULA rapidly oscillates with time scale much smaller than the Hubble time. After averaging under the WKB ansatz one can find the equation of state $\langle w_{a}\rangle=0$ and the energy density redshift as $\langle\rho_a\rangle\sim a^{-3}$, i.e., the same as CDM. 

We now turn to discuss perturbations of the ULA. The difference of ULA from CDM is the non-trivial effective sound speed, which has free-streaming behavior on small scales. We start from the perturbed equation in synchronous gauge
\begin{equation}\label{axioneom1}
    \delta\phi''+2\mathcal{H}\delta\phi'+\left(k^2+a^2m^2\right)\delta\phi=-\frac{1}{2}h'\phi',
\end{equation}
where $h$ is the trace of the scalar metric perturbation in synchronous gauge. Similarly to the background equation, $\delta\phi$ also oscillates after $\tau_m$. We can also expect the $k$-dependent term and the source term from gravity. One can also write down the  DM  fluid perturbations
\begin{align}
    \delta\rho_{a}= &\frac{1}{a^2}\phi'\delta\phi'+m^2\phi\delta\phi,\nonumber\\ 
    \delta P_{a}= &\frac{1}{a^2}\phi'\delta\phi'-m^2\phi\delta\phi, \nonumber\\
    \theta_{a}= &\frac{k^2}{\left(\rho_{a}+P_{a}\right)a^2}\phi'\delta\phi.
\end{align}
Then the equation of the ULA field perturbations can be rewritten as \cite{Hu:1998kj,Hu:2003hjx}
\begin{align}\label{axioneffeom1}
    \delta_{a}'= &-\left(1+w_{a}\right)\left(\theta_a+\frac{h'}{2}\right)-3\mathcal{H}\left(c_{s}^2-w_{a}\right)\delta_{a}\nonumber\\&-9\mathcal{H}^2\left(1+w_{a}\right)\left(c_{s}^2-c_{a}^2\right)\frac{\theta_{a}}{k^2},\nonumber\\
    \theta_{a}'= &-\mathcal{H}\left(1-3c_{s}^2\right)\theta_{a}+\frac{c_{s}^2k^2}{1+w_{a}}\delta_{a},
\end{align}
where we defined $\delta_{a}\equiv\delta\rho_{a}/\rho_{a}$.
Note that $c_{a}^2\equiv P_{a}'/\rho_{a}'$ is the adiabatic sound speed and $c_{s}^2\equiv \delta P_{a}/\delta\rho_{a}$ is the sound speed of ULA field in the comoving frame, which is given by $c_{s}^2=1$. Similar to the equation of state, the sound speed also oscillates after $\tau_m$. One can also use the WKB ansatz and average the ULA fluid after $\tau_m$. Then the effective sound speed at the leading order of $(m/H)^{-1}$ in the comoving frame of effective ULA fluid is \cite{Hwang:2009js,Park:2012ru}
\begin{equation}
    \langle c_{s}^2\rangle=\frac{k^2/\left(4m^2a^2\right)}{1+k^2/\left(4m^2a^2\right)}.
\end{equation}

Let us define the free-streaming scale $k_{m}$ as the mode which has $c_{s}^2\simeq 1$ when crossing the horizon. Namely, the physical momentum of the mode becomes relativistic
$k/a(\tau_{r})=m$ when the mode enters the horizon at $k/a(\tau_{c})=H(\tau_{c})$. Hence the $k_{m}$ is given by
\begin{equation}
    k_m=ma_{m}=H_{m}a_{m}=a_{m}\sqrt{mH_{m}},
\end{equation}
where $a_{m}=a(\tau_{m})$, $H_{m}=H(\tau_{m})=m$ and $\tau_{m}$ is the time that ULA starts to oscillate. Modes with comoving momentum $k>k_{m}$ is still free-streaming after entering the horizon hence causing the suppression of small scale structures, which is different from CDM. The suppression continues until the physical momentum $k/a$ becomes smaller than the Jeans momentum $k_{J}/a$, where
\begin{equation}\label{Jeans}
    k_{J}=a\sqrt{mH}\simeq\begin{cases}
    H_0^{1/2}\Omega_{r}^{1/4}m^{1/2},\quad & \tau<\tau_{\text{eq}},\\
    H_0^{1/2}\Omega_{m}^{1/4}a^{1/4}m^{1/2},\quad & \tau_{\text{eq}}<\tau<\tau_{0}.
    \end{cases} 
\end{equation}
This occurs after the matter-radiation equality because before $\tau_{\text{eq}}$ the physical momentum of $k$ and $k_{J}$ redshift in the same way. 
Note that $k_m$ is equal to the Jeans scale $k_{J}$ at $\tau_m$, which can be estimated as 
\begin{equation}\label{km}
    k_{m}=k_{J}(\tau_m)\simeq\begin{cases}
    H_0^{1/2}\Omega_{r}^{1/4}m^{1/2},\quad & \tau_{m}<\tau_{\text{eq}},\\
    H_0^{2/3}\Omega_{m}^{1/3}m^{1/3},\quad & \tau_{\text{eq}}<\tau_{m}<\tau_{0},
    \end{cases} 
\end{equation}
where we have used $H_{m}^2/H_0^2\simeq\Omega_{m}/a_m^3$ hence $a_{m}\simeq(\Omega_{m}H_0^2)^{1/3}m^{-2/3}$. We denote the scale $k_m$ with a subscript $``m"$ because it is mass-dependent and  different for heavy and light ULAs. In the next section we will consider multiple ULAs with different masses hence fluids become free-streaming on different scale $k_{m}$.

The free-streaming of ULA fluid in the horizon results in a fraction of total matter in the matter-dominated era. It is well-known that this can cause suppression of the total growing overdensity \cite{Arvanitaki:2009fg,Marsh:2010wq,Bond:1980ha,Amendola:2005ad}. We discuss in the synchronous gauge. There is still gauge freedom in this gauge so one have to remove it by choosing the CDM comoving frame  \cite{Ma:1995ey,Bucher:1999re}. In this choice the adiabatic CDM overdensity $\delta_{c}$ grows with metric $\delta_{c}=-h/2$. To solve the metric $h$, we can use the linearized Einstein equation to write down the closed equation of $h$
\begin{equation}
    h''+\mathcal{H}h'=-3\mathcal{H}^2\sum_{J}{\Omega_{J}\left(1+3c_{sJ}^2\right)\delta_{J}}.
\end{equation}
If CDM is the dominant heavy component to the universe in the matter-dominated era, the r.h.s of the equation is $-3\mathcal{H}^2\left(1-f\right)\delta_{c}=(3/2)\mathcal{H}^2\left(1-f\right)h$, where we have defined $f\equiv\Omega_{a}/\Omega_{m}$. Then this equation has growing solution
\begin{equation}\label{cdmoverdensity}
    \delta_{c}=-\frac{h}{2}\propto a^{p},\ \ \ \ \ \ \ \ p=\frac{-1+\sqrt{25-24f}}{4}\simeq 1-\frac{3}{5}f.
\end{equation}
The growth of matter is suppressed compared to the original total CDM case because of the free-streaming of the ULA fluid. Hence the total matter power spectrum is suppressed on small scales. 

If we consider multiple ULAs with different masses, the above discussion is still valid for each of them if there is no interactions between ULAs. To reveal the effects of single or multiple ULAs on matter power spectrum and CMB, we will numerically  solve the Einstein-Boltzmann system including multiple ULAs.

\section{Accurate Boltzmann code}
\label{sec:Boltzmanncode}
We developed our own Boltzmann code to calculate the growth of perturbations and the power spectrum in the presence of multiple ULAs. The linear matter power spectrum of $\Lambda$CDM has been well studied analytically and numerically by solving the Boltzmann system together with Einstein equations. 
In the case of ULAs, they oscillate after rolling down to the bottom of the potential. Since the oscillation has very small time scale $\sim m^{-1}$, it is difficult to calculate
the spectrum. Hence, the WKB approximation is used to average the rapid oscillation of equation of state and the sound speed of ULAs. This is what existing codes, such as {\scriptsize AXIONCAMB} \cite{Hlozek:2014lca}, have done. However, this usual effective fluid approximation for the light axion has exposed errors at the $1\sim2\sigma$ level for future experiments  \cite{Urena-Lopez:2015gur,Cookmeyer:2019rna}. Recently, an accurate effective fluid approximation for ultralight axions has been proposed \cite{Passaglia:2022bcr}, where the error is smaller and controllable. In our code, we adopt this accurate effective fluid approximation and we found it works for multiple ULAs with the mass $m\gtrsim 10^{-27}$eV.

In this paper, we consider the standard $\Lambda$CDM (including photons, massless neutrinos, baryons, CDM and dark energy) plus Multi-ULA. The treatments for recombination and mode growth of photon, baryon, massless neutrino and CDM is standard \cite{Ma:1995ey,Bucher:1999re}. Our code for these parts is based on \cite{Callin:2006qx} although we used the synchronous gauge. Effects of helium and reionization have been included in the recombination histories. Before the recombination, the photon-baryon fluid is numerically unstable  due to the large optical depth $\tau_{\text{op}}'$. To solve this problem, we expand $3\Theta_{1}+v_b$ in powers of $1/\tau_{\text{op}}'$ during tight-coupling regime as is done in \cite{Doran:2005ep}. We use the line-of-sight integration developed in \cite{Seljak:1996is} to compute the multipoles transfer functions and truncate the hierarchy of equations at $l_{\text{max}}=8$ for photons and massless neutrinos. The parameters of the $\Lambda$CDM model we use in this paper are as follows, 
\begin{align}
&h=0.674,\ \ T_{0}=2.7255\text{K},\ \ N_{\text{eff}}=3.046,\ \  \Omega_{c}=0.267,\ \ \Omega_{b}=0.05,\ \  \Omega_{k}=0,\nonumber\\ 
&Y_{p}=0.245,\ \  z_{\text{reion}}=8,\ \ \Delta z_{\text{reion}}=0.5,\ \ z_{\text{Hereion}}=3.5,\ \ \Delta z_{\text{Hereion}}=0.5.\nonumber
\end{align}

\begin{figure}[tbp]
\centering
\includegraphics[scale=0.68]{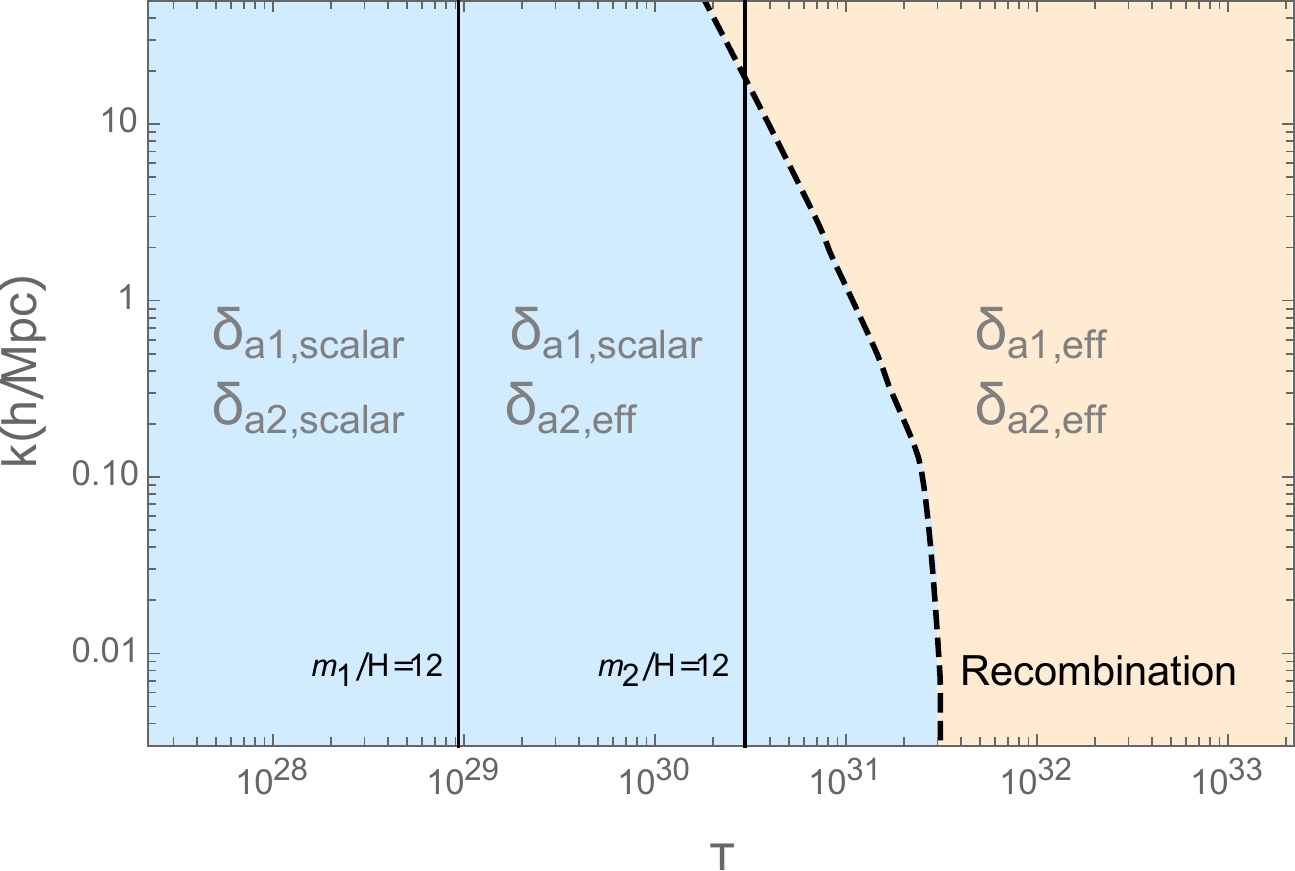}
\caption{\label{fig:ktau}The process of numerical calculation of the Boltzmann code. The transitions of scalar fields to effective fluid are chosen as $m_{1}/H=12$ and $m_{2}/H=12$ for two ULAs respectively. The blue region is the photo-electron tight-coupling regime while the orange region is the decoupling regime. The transition time from tight-coupling to the full treatment shown by a dashed curve is mode-dependent on small scales.}
\end{figure}

For the ULA parts, we start to treat the ULA field as an effective fluid from a time around $m/H_{*}$ in the oscillatory regime, which guaranteed high enough accuracy at late time for background and perturbations. Matching error
at the switching time and also the evolution error of energy density have been explored. After the choice of effective equation of state and sound speed for ULA \cite{Passaglia:2022bcr}
\begin{equation}\label{weffceff}
    w_{a,\text{eff}}=\frac{3}{2}\left(\frac{m}{H}\right)^{-2},\ \ \ \ \ c_{s,\text{eff}}^2=\langle c_{s}^2\rangle+\frac{5}{4}\left(\frac{m}{H}\right)^{-2},
\end{equation}
the evolution error is eliminated and only matching error left at late time. The error decreases as $\left(m/H_{*}\right)^{-3}$ for background $\rho_a$ but slower for perturbation $\delta\rho_{a}$ because of the scale $k$ term and the gravitational source term. Before the switching time, we directly solve the equation of motion of scalar field (\ref{axioneom}) and (\ref{axioneom1}). Then we turn to effective fluid approximation at $\tau_{*}$ with some matching conditions. After the switching time, we solve the evolution of effective ULA fluid by using (\ref{axioneffeom}) and (\ref{axioneffeom1}) with effective equation of state and sound speed (\ref{weffceff}).

We use the shooting method to obtain the initial condition of a background ULA field from the relic energy density $\Omega_{a}$. The relic energy density today can be roughly given by $\Omega_a\simeq\rho_{a}(\tau_{m})/\rho_{cri}$. Then we can estimate $\Omega_{a}$ as \cite{Marsh:2010wq}
\begin{equation}\label{shooting}
    \Omega_a\simeq\begin{cases}
\left(1/6\right)\left(9\Omega_{r}\right)^{3/4}\left(m/H_{0}\right)^{1/2}(\phi_{\text{ini}}/M_{\text{pl}})^2,\quad &\tau_m < \tau_{\text{eq}}, \\
\left(9/6\right)\Omega_{m}(\phi_{\text{ini}}/M_{\text{pl}})^2,\quad & \tau_{\text{eq}}<\tau_m<\tau_{\text{end}}.
\end{cases} 
\end{equation}
The relic density $\Omega_{a}$ can be obtained from the ULA initial condition $\phi_{\text{ini}}$, i.e., the angle $\Theta_{\text{ini}}$ of the $U(1)$ cycle when the PQ symmetry is broken. If this symmetry breaking occurs before the end of inflation, $\Theta_{\text{ini}}$ is randomly distributed on $[-\pi,\pi]$. Then we have $\phi_{\text{ini}}\in[-\pi f_{a},\pi f_{a}]$. 
We first evolve the effective ULA fluid together with other fluids from present time $\tau_0$ to switching time $\tau_{*}$ and obtain the energy density $\rho_{a*}$ at the switching time. Then we assume $\phi_{\text{ini}}=C$ and $\phi_{\text{ini}}'=0$ and start to shoot out ULA field $\phi$ from (\ref{shooting}) at initial time $\tau_{\text{ini}}$ to find the correct initial condition which shoots the desired density at the switching time. The matching conditions at the switching time can be found in Appendix \ref{IC}. For multiple ULAs, for example, two ULAs with masses $H_{\text{eq}}<m_{1}<m_{2}$ in this paper, we treat them as effective fluid individually at different switching time $m_{1}/H_{*1}=12$ and $m_{2}/H_{*2}=12$, as Figure \ref{fig:ktau} shows. 
When the lighter ULA starts to oscillate the heavier one has been redshifted as CDM, i.e., $\Omega_{\text{total-dm}}=\Omega_{c}+\Omega_{a2}$. Hence we can find the initial condition of the lighter ULA similarly to the single-ULA case. After the initial condition of lighter ULA is found, we can find the initial condition of the heavier ULA under the evolution of the lighter ULA and other fluids. 

\begin{figure}[tbp]
\centering
\includegraphics[scale=0.59]{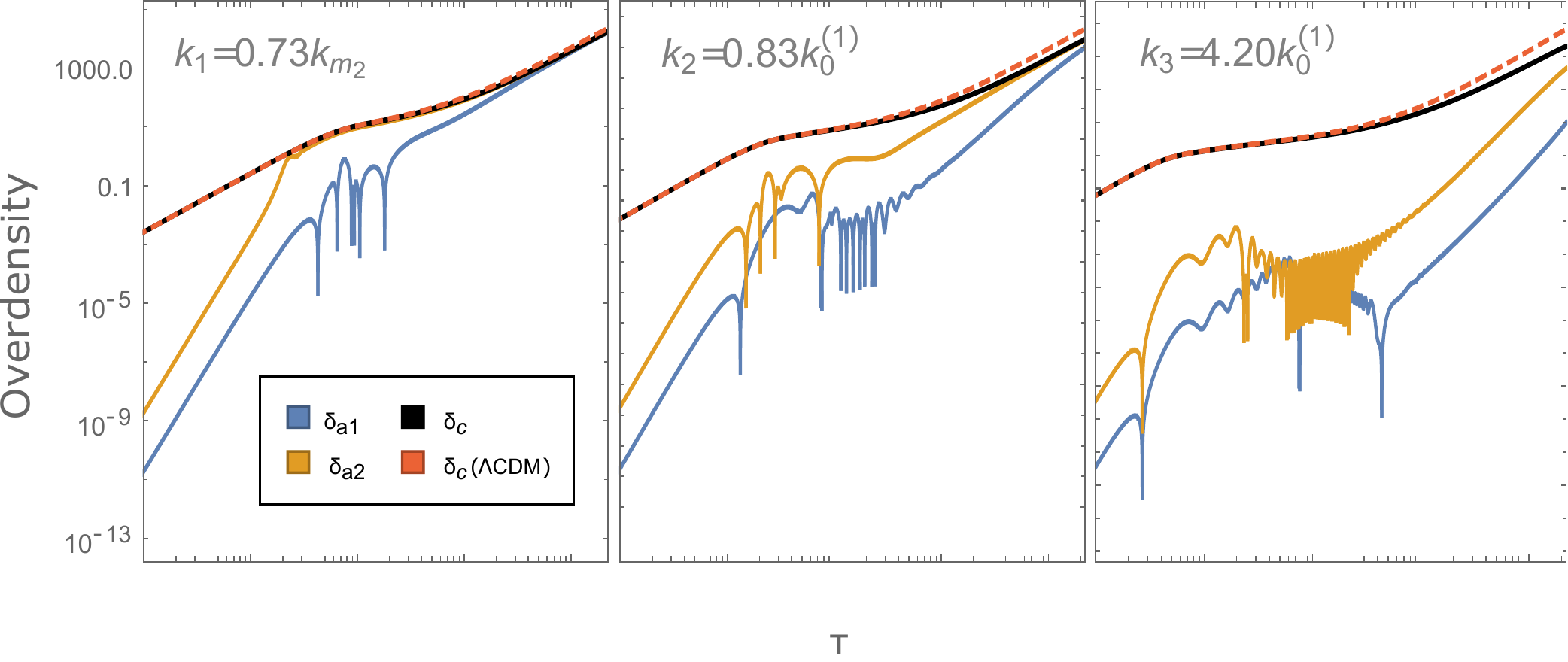}
\caption{\label{fig:overdensity}The growth of CDM and two ULAs for different three scales, $k_{m_{1}}<k<k_{m_{2}}$(Left), $k_{m_{2}}<k<k_{0}^{(1)}$(Mid) and $k>k_{0}^{(1)}$(Right). When Jeans oscillations of two ULAs are larger enough, the suppression of the growth factor is significant compared to $\Lambda$CDM. Here $m_{1}=10^{-26}$eV, $m_{2}=10^{-25}$eV and $f_{1}=0.2$, $f_{2}=0.25$.}
\end{figure}

The initial conditions of perturbations are easily determined. For adiabatic perturbations, the initial conditions $\delta\phi_{\text{ini}}$ and $\delta\phi_{\text{ini}}'$ depend on $\phi_{\text{ini}}'$, which  vanishes because of the misalignment \cite{Marsh:2010wq,Perrotta:1998vf}. Hence we simply assume all ULA perturbations are adiabatic so $\delta\phi_{\text{ini}}=0$ and $\delta\phi_{\text{ini}}'=0$ for each of them. We also switch to effective fluids of perturbations and adopt the same matching conditions at the same switch time as background fluids. Different from background, the overdensity of ULAs has Jeans oscillation on small scale hence there is also an additional matching error to $\delta_{a}$ \cite{Passaglia:2022bcr}. When considering scale under Jeans scale, this error is significant for the switching time $m/H_{*}=12$ we choose in this paper. However, if 
ULA is just a small friction of total dark matter, i.e., $\Omega_{a}/\Omega_{c}\ll1$, the overdensity $\delta_{a}$ for scale $k\gtrsim k_{m}$ will be sourced by metric perturbation and start to grow at late time, as Figure \ref{fig:overdensity} shows, hence this error is washed out\footnote{Thanks to S.  Passaglia for pointing out this.}. If $\Omega_{a}/\Omega_{c}\simeq 1$ the suppression of ULA overdensity is significant compared to CDM on Jeans scale hence this error is not important to the power spectrum anyway.

\section{Structure formation with  multi-axion dark matter}
\label{sec:Suppression}
In this section, we analytically discuss the effects of two ULAs on the total matter power spectrum. We also calculate growth of the power spectrum numerically with our code. We consider the matter power spectrum at present time
\begin{equation}\label{P}
    P(k)=\left|\frac{k^2\Psi(k)}{\left(3/2\right)\Omega_{m}H_{0}^2}\right|^2P_{\text{primordial}}(k),
\end{equation}
where $P_{\text{primordial}}(k)=\left(2\pi^2/k^3\right)A_{s}\left(k/k_{\text{pivot}}\right)^{n_{s}-1}$ is the power spectrum of primordial perturbation from inflation. We choose the amplitude $A_{s}=2\times 10^{-9}$, $n_s=0.965$ and the pivot scale $k_{\text{pivot}}=0.05\text{Mpc}^{-1}$ in our numerical calculation. $\Psi$ is the Newtonian potential and can be related to metric perturbations in synchronous gauge by gauge transformation
\begin{equation}
    \Psi=\eta-\frac{\mathcal{H}}{2k^2}\left(h'+6\eta'\right)\simeq-\frac{3}{2k^2}\mathcal{H}^2(1-f)\delta_{c},
\end{equation}
where we have used the linearized Einstein equations and assumed the CDM dominates in the matter-dominated era. Hence if the ULA fluid is free-streaming within the matter-dominated era, the suppression of $P(k)$ can be estimated from (\ref{cdmoverdensity}). 

We first discuss the single-ULA case with mass $m$ and relic density $\Omega_{a}$. Only $k>k_m$ modes can cause the suppression. The suppression starts at $a_{\text{min}}=\max{\left(a_{\text{eq}},a_{m}\right)}$.
This is because we need to consider the period $\tau_{\text{eq}}<\tau<\tau_{0}$ in (\ref{Jeans}). 
The suppression stops when $k$ becomes equal to the Jeans scale $k_J$, namely, at $a_{\text{max}}=\min{\left(a_0=1,a_{J}(k)\right)}$. Here, $a_{J}(k)$ is the scale factor when $k=k_{J}(a)$.
We have taken into account the case that $k=k_{J}(a)$ has not happen until today. 
The matter power spectrum relative to the $\Lambda\text{CDM}$ case is given by transfer function \cite{Arvanitaki:2009fg,Amendola:2005ad}
\begin{equation}
    T^2(k)\equiv\frac{P(k)}{P_{\Lambda\text{CDM}}(k)}\simeq\left(\frac{a_{\text{max}}}{a_{\text{min}}}\right)^{2(p-1)},
\end{equation}
where we used the result (\ref{cdmoverdensity}). 
Since we have the scaling $k\propto a^{1/4}$, the transfer function $T(k)$ reads
\begin{equation}
    T^2(k)=\left(\frac{k_{\text{max}}}{k_{\text{min}}}\right)^{8\left(p-1\right)},
\end{equation}
where $k_{\text{max}}=\min{(k,k_0)} $ and
$k_{\text{min}}=\max{(k_{\text{eq}},k_{m})}$. 
Here, we defined the Jeans scale at present
\begin{equation}
    k_{0}\equiv k_{J}(a_{0})= H_0^{1/2}\Omega_{m}^{1/4}m^{1/2},
\end{equation}
The suppression occurs below the Jeans scale $k_J$ which increases after $\tau_{\text{eq}}$. The relevant scales to determine a step-like feature in the matter power spectrum are $k_{\text{min}}$ and  $k_{\text{max}}$. The point is that the suppression of the mode $k$ will stop at the time $a_J(k)$ where the Jeans scale becomes larger than $k$. If $k>k_0$, this never occurs. Note that these scales are determined by the ULA mass. While the power of suppression is determined by the relative relic density $f =\Omega_{a}/\Omega_{m}$.

We consider two ULAs in order to reveal the effects of multi-axion dark matter on the matter spectrum. 
A light ULA-1 has a mass $m_{1}$ and a relic density $\Omega_{a1}$ and a heavy ULA-2 has a mass $m_{2}>m_{1}$ and a relic density $\Omega_{a2}$.
For the analysis, it is convenient to define 
\begin{equation}
    f_{1}\equiv \frac{\Omega_{a1}}{\Omega_{m}},\ \ \ \ \ \ \ f_{2}\equiv \frac{\Omega_{a2}}{\Omega_{m}},\ \ \ \ \ \ \
    \omega_{r}\equiv \frac{\Omega_{a1}}{\Omega_{a2}}.
\end{equation}
We study two ULAs oscillating before the matter-radiation equality, 
namely,  $k_{m} > k_{\text{eq}}  $. Thus,
the suppression structure is determined by the interval $k_{m_{i}} \leq k \leq k^{(i)}_0 $ for each ULA-$i$, ($i=1,2$).
There are two different scenarios for two-ULA case: the intervals of suppression have an overlap or not. Because the light ULA-1 suppresses the power spectrum on larger scale, there is an overlap if $k_{0}^{(1)}>k_{m_{2}}$. The critical case is given by
\begin{equation}
    m_{r}\equiv\frac{m_{1}}{m_{2}}=\sqrt{\frac{\Omega_{r}}{\Omega_{m}}}\simeq 0.017.
\end{equation}
When $m_{r}>0.017$, there is an overlap of two suppressed intervals. In the overlapped region, both two ULAs suppress the power spectrum on these scales. We discuss these two cases separately. 

\subsection{\texorpdfstring{Two-ULA: Cases of $m_{r}\gtrsim0.017$}%
{Something with beta in it}}

For $k<k_{m_{1}}$, ULAs behave like ordinary CDM because they have become non-relativistic when they enter the horizon. For $k_{m_{1}}<k<k_{m_{2}}$, the modes of ULA-1 becomes free-streaming while  ULA-2 behaves as  CDM, as is seen in Figure \ref{fig:overdensity}(Left). Hence, in this interval, the transfer function   is the same as that of the single ULA,
\begin{equation}\label{T1}
     \mathop{T^2(k)}\limits_{k_{m_{1}}<k<k_{m_{2}}}=\left(\frac{k}{k_{m_{1}}}\right)^{8(p_{1}-1)}
\end{equation}

\begin{figure}[tbp]
\centering
\includegraphics[scale=0.53]{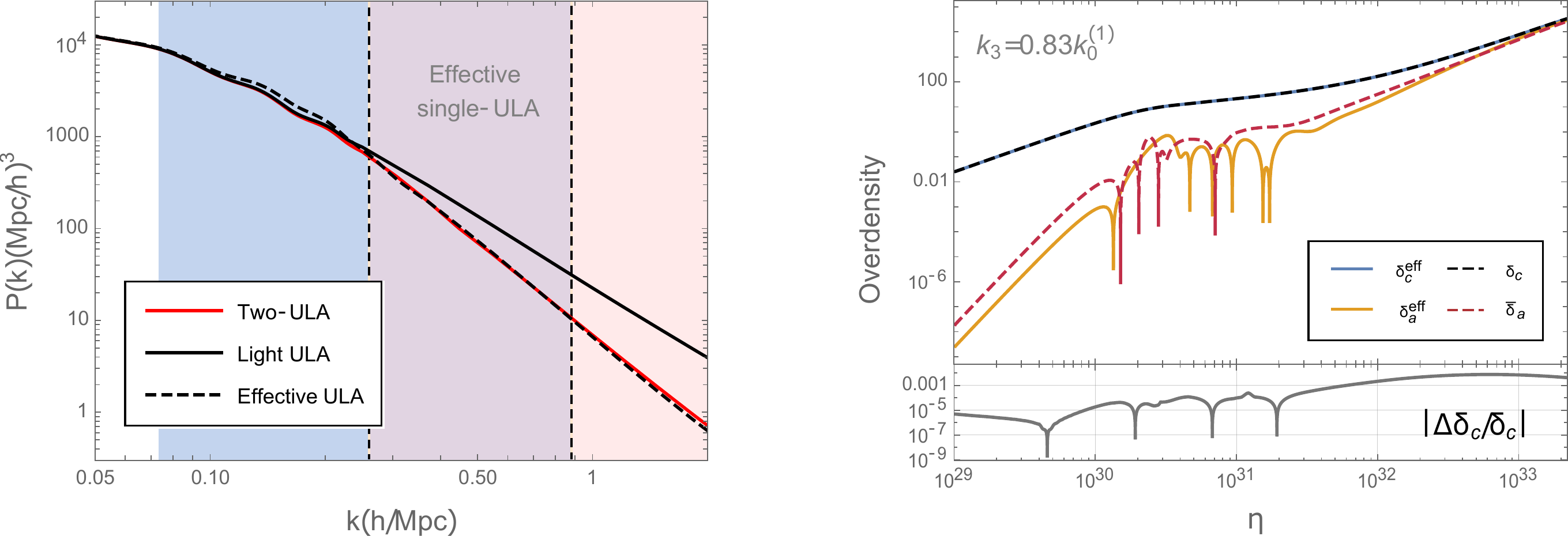}
\caption{\label{fig:EFO}(Left) The matter power spectrum of two ULAs (red curve). The purple region is the overlap of suppression of two ULAs. (Right) The growth of combined overdensity $\bar{\delta}_{c}$ and overdensity of effective single-ULA $\delta_{c}^{\text{eff}}$ on scale of overlap region $k_{m_{2}}<k<k_{0}^{(1)}$. The error of CDM is $\lesssim 10^{-3}$ during the period. We took $m_{1}=10^{-26}$eV, $m_{2}=10^{-25}$eV and $f_{1}=0.1$, $f_{2}=0.15$.}
\end{figure}

When $k_{m_{2}}<k<k_{0}^{(1)}$,  both of them are free-streaming until their physical momentum become smaller than the Jeans momentum, as you can see in Figure \ref{fig:overdensity}(Mid). When modes with comoving momentum $k$ enter the matter-dominated era, ULA-1 and ULA-2 are both free-streaming until $a_{J}^{(2)}$. 
Next during $a_{J}^{(2)}$ and $a_{J}^{(1)}$ only ULA-1 is free-streaming. 
Here, we should recall the relation $k^4= H_0^4 \Omega_m a_J m^2$.
From Eq.(\ref{km}),  we obtain
$k_{m}=H_0^{1/2}\Omega_r^{1/4}m^{1/2}$.
Using these relations, we can deduce the transfer function for this interval as
\begin{equation}\label{Tefo}
    \mathop{T^2(k)}\limits_{k_{m_{2}}<k<k_{0}^{(1)}}=\left(\frac{a_{J}^{(2)}(k)}{a_{\text{eq}}}\right)^{2(p_{12}-1)}\left(\frac{a_{J}^{(1)}(k)}{a_{J}^{(2)}(k)}\right)^{2(p_{1}-1)}=\left(\frac{k}{k_{m_{e}}}\right)^{8(p_{e}-1)},
\end{equation}
where we used $p_{12}\equiv p(f_{12})$, $f_{12}\equiv f_{1}+f_{2}$ and $p_{1}\equiv p(f_{1})$. We also defined  an effective mass $m_{e}$ and $p_{e}$ as
\begin{align}
   & m_{e}\simeq m_{1} \left(\frac{1}{m_{r}}\right)^{\frac{1}{1+\omega_{r}}},\label{Meff}\\ &p_{e}\equiv p_{12}\simeq 1-\frac{3}{5}f_{12}
\end{align}
for $f_{1},f_{2}\ll 1$. Because $p_{12}\leq p_{1}\leq 1$, we see that 
\begin{equation}
    m_{1}\leq m_{e}\leq m_{2} \ .
\end{equation}
 Hence, we have $k_{m_{1}}\leq k_{m_{e}}\leq k_{m_{2}}$ and $k^{(1)}_{0}\leq k^{(e)}_{0}\leq k^{(2)}_{0}$. That means if we have only one ULA with mass $m_{e}$, the suppression interval given by $k_{m_{e}}<k<k^{(e)}_{0}$ covers the overlap interval we discussed in (\ref{Tefo}). Hence the suppression of the power spectrum of two ULAs in overlap $k_{m_{2}}<k<k_{0}^{(1)}$ can be reproduced by an effective ULA with the moderate mass $m_{e}$ and the effective relic density parameter $\Omega_{m_{e}}=\Omega_{a1}+\Omega_{a2}$. As we see in Figure \ref{fig:EFO}(Left), when the structure suppression starts, the power spectrum follows that of ULA-1. When reaching scale $k_{m_{2}}$, where ULA-2 becomes relevant, the power spectrum coincides with that of the effective single-ULA. In Figure \ref{fig:EFO}(Right), we can see the evolution of overdensity of the effective single-ULA is similar to that of the combination of two ULAs at late time, where the combined overdensity $\bar{\delta}_{a}$ is defined as
\begin{equation}
    \bar{\delta}_{a}\equiv \frac{\rho_{a1}\delta_{a1}+\rho_{a2}\delta_{a2}}{\rho_{a1}+\rho_{a2}}.
\end{equation}
It is shown that the $\bar{\delta}_{a}$ starts to grow earlier than the effective one $\delta_{a}^{\text{eff}}$ because $\delta_{a2}$ has experienced the Jeans instability and starts to grow earlier than $\delta_{a}^{\text{eff}}$. While, when effective ULA $\delta_{a}^{\text{eff}}$ starts to grow, $\delta_{a1}$ is not in the regime of Jeans instability. Hence the growth of $\delta_{a}^{\text{eff}}$ is faster than  $\bar{\delta}_{a}$. At late time, $\bar{\delta}_{a}$ and $\delta_{a}^{\text{eff}}$ show similar growth because $\delta_{a1}$ also starts to grow and contribute to the total DM. The difference of the CDM is small hence results in the same suppression of the power spectrum at any redshift.

Now we discuss the suppression on the smaller scales $k>k_{0}^{(1)}$. On this scales,  the modes $k$ of ULA-1 never exceeds the Jeans scale. While, ULA-2 contribute to the suppression until $a_{\text{max}}^{(2)}=a_J^{(2)}(k)$. Hence the transfer function can be written as 
\begin{equation}\label{lightT}
    \mathop{T^2(k)}\limits_{k>k_{0}^{(1)}}=\left(\frac{a_{\text{max}}^{(2)}}{a_{\text{eq}}}\right)^{2(p_{12}-1)}\left(\frac{1}{a_{\text{max}}^{(2)}}\right)^{2(p_{1}-1)}=\left(\frac{k_{\text{max}}^{(2)}}{k_{m_{n}}}\right)^{8(p_{n}-1)},
\end{equation}
where $k_{m_{n}}$ is the scale given by (\ref{km}) with the following mass $m_{n}$ and the parameter $p_{n}$ 
\begin{align}
   &m_{n}\simeq m_{2}\left(0.017\right)^{\omega_{r}}\label{lightM} \ ,\\
   &p_{n}\equiv p_{12}-p_{1}+1 \simeq p_2 \ .\label{lightp}
\end{align}
Here, we indicated that the parameter $p_{n}$ is not exactly equal to $p_{2}$ but smaller $p_{n}<p_{2}$. This could be understood as follows. When ULA-2 becomes free-streaming, the ULA-1 has suppressed the growth of dark matter for a while. Hence the potential well provided by matter has discounted and ULA-2 is more difficult to grow than only one ULA case. But for $f_{1},f_{2}\ll 1$ this effect is small $p_{n}\simeq1-3f_{2}/5$. Note that here we define the $m_n$ just for convenience. It does not mean the suppression can be reproduced by single ULA with effective mass because $m_{n}<m_{2}$. One can not find one ULA with mass lighter than $m_{2}$ or has small relic $\Omega_{a2}$ but still suppressing the power spectrum on such small scale. In other words, the de Broglie wavelength of ULA is so long that it is difficult to detect its variation on such a small scale until today. 

\begin{figure}[t]
\centering
\includegraphics[scale=0.53]{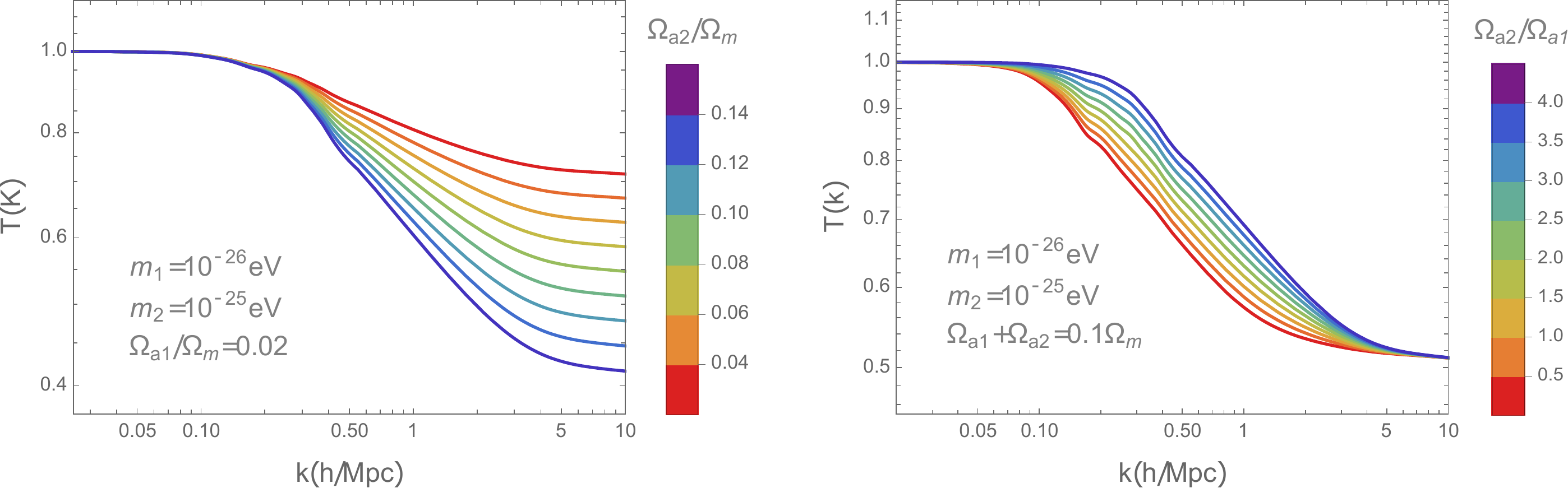}
\caption{\label{fig:Omega}(Left) The transfer function for different relic $\Omega_{2}$ with fixed $\Omega_{1}=0.02\Omega_{m}$. (Right) The transfer function for  different relative relic $\omega_{r}$ with fixed $\Omega_{a1}+\Omega_{a2}$. When we increase $\omega_{r}$, the transfer function $T(k)$ approaches that of ULA-1.}
\end{figure}

The effective mass $m_{e}$ depends on the mass of two ULAs and their relic energy densities. If we fix $m_{1}$ and $m_{2}$, the effective mass exponentially depends on  $(1+\omega_{r})^{-1}$. This is because the growth of overdensity is exponentially suppressed by the fraction of free-streaming fluids. Therefore increasing $\Omega_{a2}$ increases the total amount hence weights the suppression exponentially, as Figure \ref{fig:Omega}(Left) shows. If we fix $\Omega_{a1}+\Omega_{a2}$ and add more amount of the heavy ULA, the free-streaming of the heavier ULA becomes more important. Hence the power spectrum is more close to that of the heavier ULA, as Figure \ref{fig:Omega}(Right) shows.

\subsection{Two-ULA: Cases of \texorpdfstring{$m_{r}\lesssim0.017$}%
{Something with beta in it. }}
\begin{figure}[t]
\centering
\includegraphics[scale=0.5]{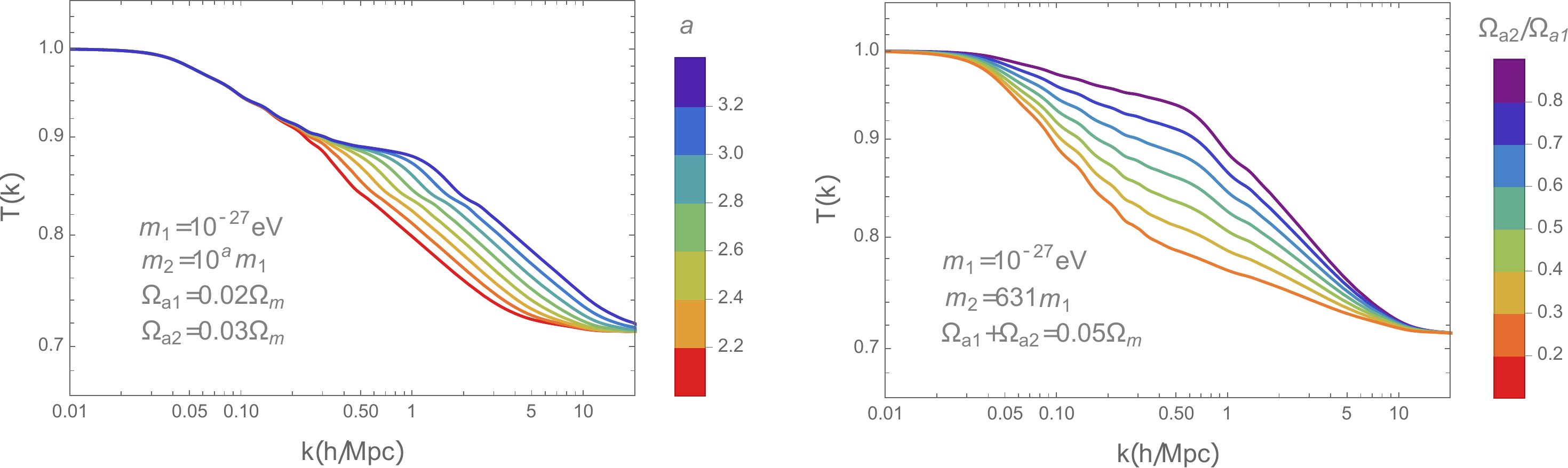}
\caption{\label{fig:OmegaO}(Left) Case $m_{r}>0.017$: the transfer function for different mass $m_{2}$ with fixed $m_{1}$, $\Omega_{a1}$ and $\Omega_{a2}$. Decreasing $m_{r}$ separate the effects of two ULAs so the multiple steep-like feature occurs. (Right) The mass hierarchy $m_{r}\ll 0.017$: The transfer function for different relative relic $\omega_{r}$ with fixed masses and total relic.  The suppression of ULA-1 and ULA-2 are independent.}
\end{figure}

In cases $m_{r}\lesssim0.017$, the inequality $k_{0}^{(1)}<k_{m_{2}}$ holds.  Hence, the relevant intervals to the suppression of ULA-1 and ULA-2 are separated. There are two cases to be considered: $k_{m_{1}}<k<k_{m_{2}}$ and $k>k_{m_{2}}$. In the former case, the suppression of the power spectrum is provided only by the ULA-1. While, in the latter case,  the ULA-2 can also contribute to the suppression. The transfer function of the former one is the same as single-ULA case, i.e.,
\begin{equation}
     \mathop{T^2(k)}\limits_{k_{m_{1}}<k<k_{m_{2}}}=\left(\frac{k_{\text{max}}^{(1)}}{k_{m_{1}}}\right)^{8(p_{1}-1)}.
\end{equation}
For the transfer function of the later case, the suppression of ULA-2 will stop at $a_J^{(2)} (k)$. Hence, we have 
\begin{equation}
    \mathop{T^2(k)}\limits_{k>k_{m_{2}}}=\left(\frac{k_{\text{max}}^{(2)}}{k_{m_{n}}}\right)^{8(p_{n}-1)},
\end{equation}
where $m_{n}$ and $p_{n}$ are given (\ref{lightM}) and (\ref{lightp}). Also, $m_{n}$ is not an effective mass of single ULA because two ULAs are almost independent. This situation is more simple and we can see a multiple steep-like feature in the transfer function if increasing the mass hierarchy. As Figure \ref{fig:OmegaO}(Left) shows, the multiple steep-like feature becomes visible when $m_{2}/m_{1}\gtrsim \mathcal{O}(100)$. For fixed total amounts of ULA density, the separation of suppression of two ULAs is clear if their relic density contrast is larger enough, as Figure \ref{fig:OmegaO}(Right) shows. Therefore we can just treat the effects of two ULAs separately on different scales.


We make some comments to the suppression of two ULAs. Firstly, we found an effective single-ULA replacement for the overlap suppression of two ULAs, which scales as $\ln{m_{e}}\propto (1+\omega_{r})^{-1}$. This is a good characteristic mass scale to the two ULAs because one can roughly estimate the half-mode scale of transfer function, even when all DM is composed of ULAs. We will discuss it in Section \ref{sec:WDM}. Secondly, one of the difference from single-ULA is that it extends the interval of suppression.  If  $m_{r}\gtrsim0.017$, the slope of the power spectrum in the interval are different on three different scale intervals: $\Omega_{a1}$ for $k_{m_1}<k<k_{m_2}$, $\Omega_{a1}+\Omega_{a2}$ for $k_{m_{2}}<k<k_{0}^{(1)}$ and $\Omega_{a2}$ for $k_{0}^{(1)}<k<k_{0}^{(2)}$. The slope is gentler in the first and thrid intervals. Hence if we keep the total amounts of ULA relic density unchanged, which means the step size of transfer function remains unchanged, extending the suppression interval makes the average slope of suppression gentler. This is more significant when all DM is composed of ULAs, as we will see shortly. Hence the transfer function is significantly distinguished from one of the single-ULA case. The Lyman-$\alpha$ forest constraint is sensitive to the shape of the power spectrum since the mass of dark matter haloes scales as radius cubed. The difference of transfer function will cause significant change to the abundance of haloes, as we will discuss in the next section.

\subsection{Multi-ULA cases}
\label{sec:Multi}

So far, we have investigated two ULAs in detail.
Now, we can infer more general results. 
Let us consider N-ULAs with mass $m_i (i=1,\cdots , N)$.
If suppression from Heaviest axion (with mass $m_h$) and the lightest axion (with mass $m_l$) has overlap, i,e, $k_{m_h}<k_{0}^{(m_l)}$. Then in this overlap interval of $k$ all axions suppress the structure. 
The condition of overlap is universal, namely, the critical 
ratio $m_r = m_i /m_j =0.017$ is the criterion.
In the case of two ULAs, we observed that two ULAs can be described by the effective single ULA with the effective mass.
First of all, we notice that the effective mass in (\ref{Meff})
can be written as 
\begin{align}
m_{e}= m_{1}^{\frac{\Omega_{a1}}{\Omega_{a1} +\Omega_{a2}}} m_{2}^{\frac{\Omega_{a2}}{\Omega_{a1} +\Omega_{a2}}}  \ .
\end{align}
Hence, we infer the general rule for the case there are overlap among $N$-ULAs as follows
\begin{eqnarray}
  \log m_e = \sum_i^{N} N_i \log m_i \ ,
\end{eqnarray}
where $N_i$ is the fraction of $i$-th axion $N_i = \Omega_i/\sum_i^N \Omega_i$. The effective axion has
an abundance $\Omega_e = \sum_i^N \Omega_i$.
Thus, once the mass distribution of multi-ULA is given,
we can analytically infer the power spectrum.

\subsection{Comparison with WDM}
\label{sec:WDM}

It is known that WDM can also suppress the power spectrum on small scales. The difference is that  WDM particles are thermally produced by decoupling from cosmic plasma while it is still relativistic. Although the light particles become non-relativistic before the matter-radiation equality, they still have large sound speed and freely move in the universe. Therefore the small scale structure is washed out and the power spectrum is suppressed on small scales. 

In the case where all dark matter is composed of WDM, the transfer function of WDM can be parametrized as \cite{Bode:2000gq}
\begin{equation}\label{TWDM}
    T_{\text{WDM}}(k)=\left(1+(\alpha k)^{2\mu}\right)^{-5/\mu},
\end{equation}
where $\mu=1.12$, $\alpha=0.074(m_{X}/\text{keV})^{-1.15}(h/0.7)$Mpc and $m_{X}$ is the thermal relic mass of WDM particles. On the other hand, in the case where all DM is single-ULA, we adopt the improved empirical form of the transfer function in \cite{Passaglia:2022bcr}, 
\begin{equation}\label{TULA}
    T_{\text{ULA}}(k)=\frac{\sin{(Ak/k_{a})^n}}{\left(Ak/k_{a}\right)^n\left[1+B \left(Ak/k_{a}\right)^{6-n}\right]},
\end{equation}
where $k_{a}= 9 m_{22}^{1/2}$, $A= 2.22m_{22}^{1/25-1/1000\ln{(m/10^{-22})}}$, $B=0.16m_{22}^{-1/20}$, $m_{22}\equiv m/10^{-22}$ and $n=5/2$. The half-mode scale $k_{1/2}$ is defined by the following equation
\begin{equation}
    T(k_{1/2})=\frac{1}{2}\left(1-T(k\to \infty)\right).
\end{equation}
To compare the ULA and WDM, we choose to match the half-mode of their transfer function $T_{\text{ULA}}(k_{1/2})=T_{\text{WDM}}(k_{1/2})=0.5$. From (\ref{TWDM}) and (\ref{TULA}), one can find the correspondence
\begin{equation}\label{mX}
    m_{X}=0.82m_{22}^{0.39}\ \text{keV}.
\end{equation}
For example, the ULA with mass $m= 10^{-23}$eV, which suppresses the power spectrum on the scale $k_{1/2}\sim 2 \text{Mpc}^{-1}$, corresponds to WDM with mass $m_{X}\sim0.3$keV. 

\begin{figure}[tbp]
\centering
\includegraphics[scale=0.58]{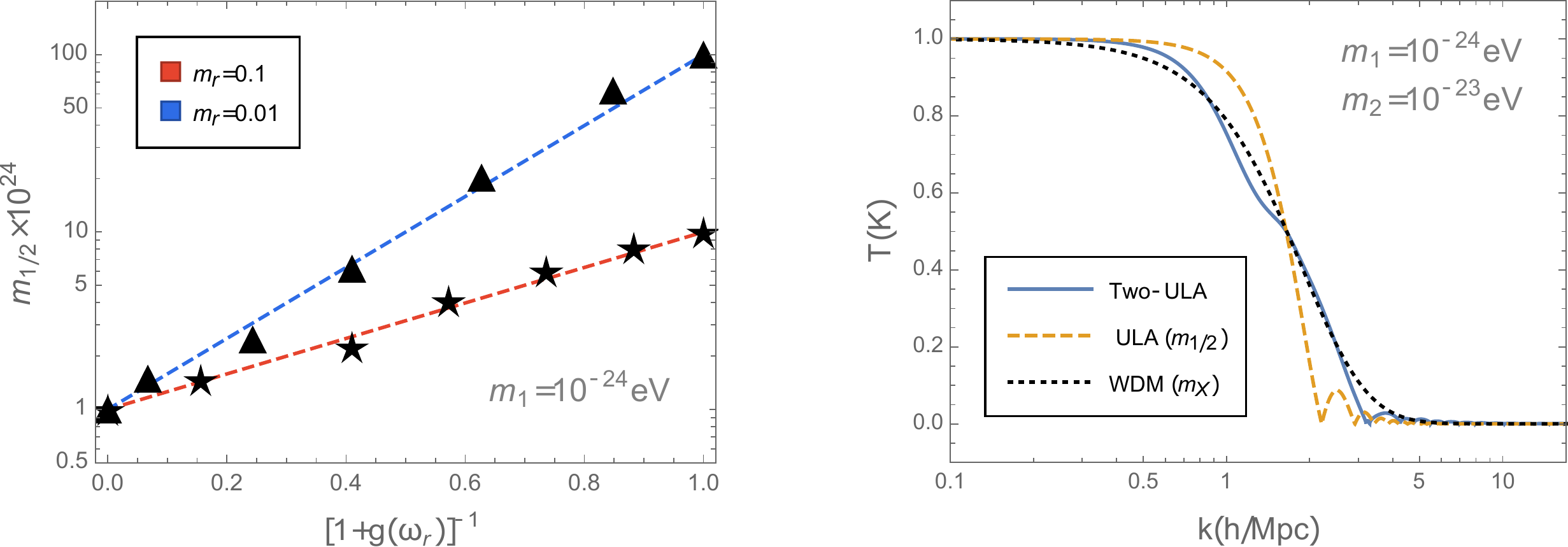}
\caption{\label{fig:m12}(Left) The numerical results of half-mode mass $m_{1/2}$ v.s. $(1+g(\omega_{r}))^{-1}$(black stars and black triangles) and their numerical fits (\ref{m1/2})(dashed curves). (Right) The transfer functions for two-ULAs,  single-ULA with $m_{1/2}$ and WDM with mass $m_{X}$(the third star in the left figure), which share the same $k_{1/2}$. The slop of the power spectrum for two ULAs is not as steep as that of the single-ULA one.}
\end{figure}

We next focus on comparison of the transfer function of multiple ULAs to the single-ULA and WDM. In the multiple ULAs, if all DM is composed of ULAs, the power spectrum is significantly suppressed on Jeans scale $k_{m2}$ and we have no  formula of the transfer function for this case hence we can not find the half-mode scale directly. However, we found that for $f_{1},f_{2}\ll 1$ and $m_{r}\gtrsim0.017$, the half-mode scale of the single-ULA with effective mass (\ref{Meff}) is a good characteristic scale to the two ULAs case. This is because, if $\Omega_{a1}\sim\Omega_{a2}$, the half-mode scale $k_{1/2}$ is in the effective single-ULA scale interval $(k_{m_{1}},k_{m_{2}})$. In other words, the two ULAs and a single ULA with the effective mass have the same half-mode scale. If $\omega_{r}\ll1$($\omega_{r}\gg1$), although the 
half-mode scale stay outsider the effective ULA interval, the effective mass is close to $m_{2}$($m_{1}$) and the transfer function of two ULAs is similar to that of $m_{2}$($m_{1}$). Therefore we can also use the half-mode scale of the effective single-ULA as that of two ULAs.

We motivated by the discussion above and the exponential sensitivity of the effective mass to $(1+\omega_{r})^{-1}$ in (\ref{Meff}). We numerically found the half-mode scale of two ULAs can be roughly fitted by the half-mode scale of single-ULA with the mass
\begin{align}\label{m1/2}
    m_{1/2}=m_{1}\left(\frac{1}{m_{r}}\right)^{\frac{1}{1+g(\omega_{r})}},\qquad
    g(x)=x+0.55\sqrt{\frac{1}{m_{r}}}x^2,
\end{align}
as is shown in Figure \ref{fig:m12}(Left). The dashed curve is the estimation using (\ref{m1/2}), while black stars stand for the mass of single-ULA which sharing the same half-mode scale with two ULAs in our numerical calculation, see the Figure \ref{fig:m12}(Right). For $m_{r}=0.1$ and $\omega_{r}\gtrsim 0.5$, we can see the mass $m_{1/2}$ approaches that of light ULA quickly. For larger mass hierarchy, this occurs at smaller $\omega_{r}$. This is because the factor $\sqrt{1/m_{r}}$ in $g(\omega_{r})$. If the mass hierarchy of two ULAs is large enough, they suppress on very different scales. So when $\omega_{r}$ becomes large enough, the half-mode scale suddenly changes from ULA-2 to ULA-1. Hence, we describe this transition by this exponential form.

\begin{figure}[tbp]
\centering
\includegraphics[scale=0.6]{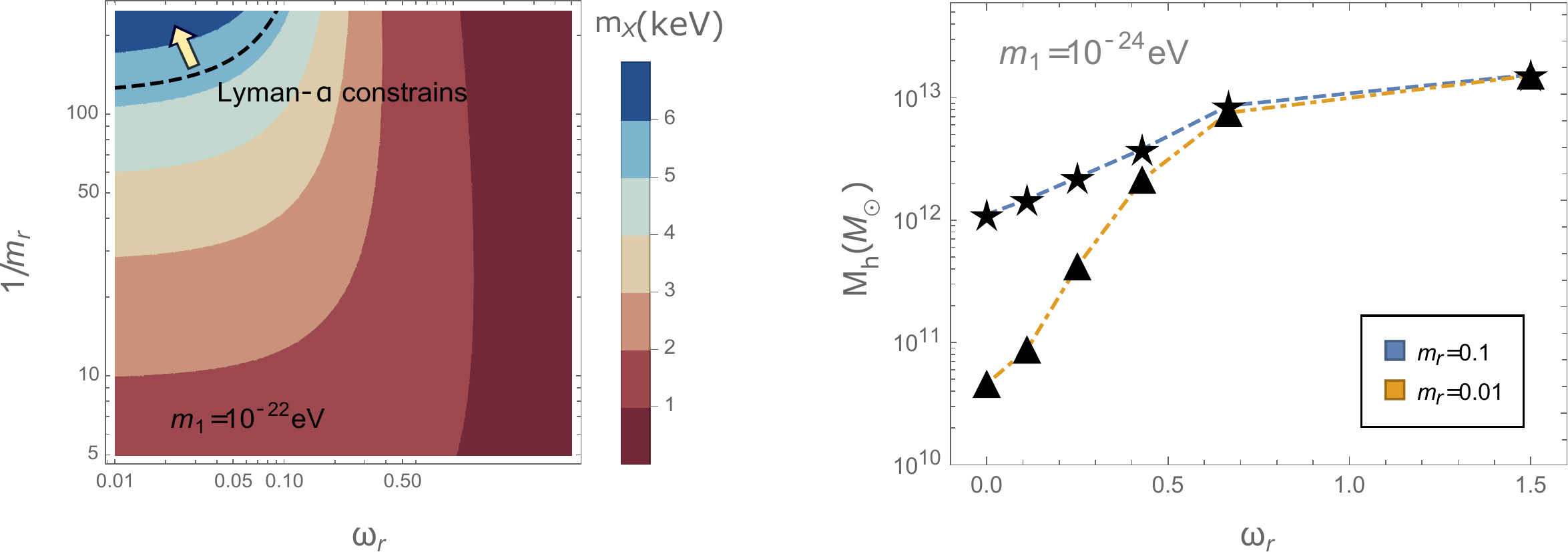}
\caption{\label{fig:6d3t}(Left) The parameter space of WDM mass $m_{X}$ v.s. $m_{r}$ and $\omega_{r}$ for $m_{1}=10^{-22}$eV. We plot until $1/m_{r}<250$. For larger $m_{X}$, the relic of heavier ULA should be dominated. This opens only a small window in the upper left from Lyman-$\alpha$ constraints on WDM mass $m_{X}>5.3$keV. (Right) The characteristic mass of the scale $k_{1/2}$. The mass of $\omega_{r}=0$ is low by a few of orders of magnitude than the one of $\omega_{r}\gtrsim 1$.}
\end{figure}

By replacing the half-mode scale of two ULAs with that of the single-ULA with mass $m_{1/2}$, we can compare the two-ULA with the WDM. From (\ref{mX}), we have the relation
\begin{equation}\label{mxm1m2}
    m_{X}=0.82 \left(\frac{m_{1}}{10^{-22}\text{eV}}\right)^{0.39}\left(\frac{1}{m_{r}}\right)^{\frac{0.39}{1+g(\omega_{r})}}\ \text{keV}.
\end{equation}
We found that this estimation works until the mass hierarchy $m_{r}\sim0.004$. We can see in the Figure \ref{fig:m12}(Right), The average slope of the transfer function of two ULAs is smaller than that of corresponding single-ULA and close to that of WDM. This is due to the extension of suppressing interval of two ULAs. Hence the shape of power spectrum of two ULAs and single ULA can be quite distinguishable from observations. Here we consider a fixed light ULA $m_{1}= 10^{-22}$eV. From the estimation (\ref{mxm1m2}), we can plot the parameter space of different mass hierarchy $1/m_{r}$ and relative relic $\omega_r$ for $m_{1}=10^{-22}$eV, which is the mass of solving cusp-core issue of halo core, as Figure \ref{fig:6d3t}(Left) shows. The mass of WDM particles is constrained as $m_{X}>5.3$keV from Lyman-$\alpha$ forest \cite{Irsic:2017ixq}, which corresponds to the small window  $100\lesssim m_r\lesssim 250$ and $\omega_r\lesssim0.05$. We can see ULAs are dominated by the heavy one. However, the Lyman-$\alpha$ constraint is sensitive to the shape of power spectrum and we have seen different $\omega_r$ can significantly change this shape. This shape can be also distinguishable from WDM so this comparison is somehow crude. For larger mass hierarchy, the suppression of two ULAs are on separate scales. In this case, half-mode scale may not be a good characteristic scale for the suppression. We will discuss it in the next section.

We can also associate characteristic masses of a sphere of radius $R=\pi/k$ to the half-mode scales $k_{1/2}$
\begin{equation}
    M_{h}=\frac{4}{3}\pi\left(\frac{\pi}{k_{1/2}}\right)^3\rho_{m},
\end{equation}
where $\rho_{m}=\rho_{\text{cr}}\Omega_{m}$ is the matter density of our universe. We expect that the formation of haloes below $M_{h}$ is suppressed significantly due to the suppression of linear power spectrum. We learned the dependence of half-mode scale on different relative relic $\omega_{r}$ and $m_{r}$ for ULAs with mass $m_{1}=10^{-24}$eV in Figure \ref{fig:6d3t}(Right). We have found that the shape of power spectrum of two ULAs is deformed for $\omega_{r}\lesssim1$. That means the half-mode scales is more sensitive to $\omega_{r}$ in this range. For $m_r=0.1$, the characteristic mass $M_h$ varies more than one order of magnitude between $0<\omega_{r}<1$, see Figure \ref{fig:6d3t}(Right) . This difference of magnitude becomes two order for $m_{r}=0.01$. For $\omega_{r}>1$, the variance of $M_h$ is insignificant, and also so for different mass hierarchies. Therefore, the structure formation is needed to be carefully explored for multiple ULAs.

\section{Abundance of dark matter haloes}
\label{sec:haloes}
The linear matter power spectrum on small sales is modified by the effects of multiple ULAs hence the mass spectrum of dark haloes should be different from that of the $\Lambda$CDM. Although the dark matter halo collapse is a non-linear process, the mass spectrum of haloes can be modeled with the linear overdensity~\cite{Press:1973iz}. The linear power spectrum offers the halo mass function, which is the number density of the collapsed dark matter haloes at specific redshift per unit mass per unit volume
\begin{equation}
    \frac{d n}{d\ln{M}}=-\frac{1}{2}\frac{\rho_{m}}{M\sigma^2}f(\nu)\frac{d\sigma^2}{d\ln{M}},
\end{equation}
where $M=(4/3)\pi R^3\rho_{m}$ is the mass enclosed within a sphere of radius $R$. Here, $f(\nu)$ is the first-crossing probability  for the ellipsoidal collapse \cite{Sheth:1999mn}
\begin{equation}
    f_{\text{fc}}(\nu)=A\sqrt{\frac{2q\nu}{\pi}}\left[1+\left(q\nu\right)^{-p}\right]e^{-q\nu/2}, \ \ \ \ \ \ \  \nu=\left(\frac{\delta_{\text{crit}}}{\sigma}\right)^2,
\end{equation}
where $A=0.3222$, $p=0.3$, $q=0.707$ and $\delta_{\text{crit}}$ is the critical collapse linear overdensity. In the Press-Schechter model, the feature of the mass function is characterized by the variance $\sigma^2$ of the matter overdensity within a given size $R$, which can be computed from the linear power spectrum as \footnote{In this section, we rescale the power spectrum $P(k)\to\Omega_{m}^2P(k)$ because (\ref{P}) is the correlation function of $\delta_m/\Omega_{m}$}
\begin{equation}\label{sigma2}
    \sigma^2(R)=\int_{0}^{\infty}{dk} k^2P(k)W^2(kR) \ .
\end{equation}
Here, we used the top-hat window function
\begin{equation}
    W(kR)=3\frac{j_{1}(kR)}{kR}=3\frac{\sin{(kR)-(kR)\cos{(kR)}}}{(kR)^3},
\end{equation}
where $j_{l}$ is the spherical Bessel function. There are some other window functions for different DM models in order to achieve the correct predictions under some cut-off scales. We here use the top-hat filter and a scale-dependent critical density for collapsed objects, as we will see in the last subsection.

\subsection{\texorpdfstring{Variance $\sigma^2$}%
{Something with beta in it}}
\label{sec:sigma}

\begin{figure}[tbp]
\centering
\includegraphics[scale=0.58]{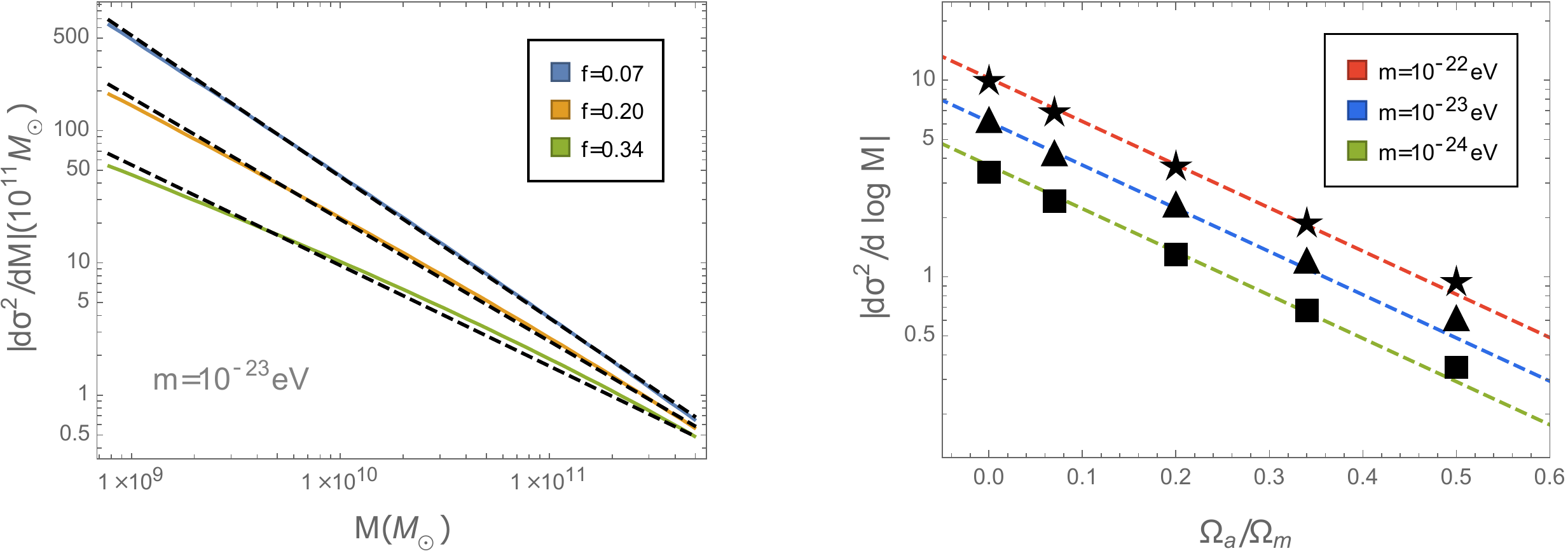}
\caption{\label{fig:Sig}(Left)The plots of derivative $d\sigma^2/dM$ of single-ULA for different fractions. The solid curves are the numerical results while the dashed curves are fitted by (\ref{dsigmaFit}). (Right)The derivative of variance for single-ULA v.s. $\Omega_{a}/\Omega_{m}$, at $M=5.62\times10^8M_{\odot}$ for $m=10^{-22}$eV,  $M=1.78\times10^{10}M_{\odot}$ for $m=10^{-23}$eV and $M=5.62\times10^{11}M_{\odot}$ for $m=10^{-24}$eV. They both scale as $\ln{(|d\sigma^2/\ln{M}|)}\propto \Omega_{a}/\Omega_{m}$. The black stars, triangles and squares are numerical results while the dashed curves are fitted by (\ref{dsigmaFit}).}
\end{figure}

One can analytically compute the variance at present time under some approximations. We are interested in the modes entering horizon before the matter-radiation equality hence we have $k>\tau_{\text{eq}}^{-1}$. On the other hand, we are considering the structure formation for spherical size $R$ hence the integral is saturated at $k<\pi/R$. This is what spherical Bessel function $j_{1}(kR)$ 
did in (\ref{sigma2}): after $kR\sim \pi$ function $j_{1}(kR)$ drops down to zero. For $kR<\pi$ the window function $W^2(kR)$ changes slowly with $\log{(kR)}$ hence we can set $W^2\simeq1$ in the integral. For $\Lambda$CDM model, it is known that the power spectrum on small scales ($k\tau_{\text{eq}}\gg 1$) is given by \cite{Gorbunov:2011}
\begin{equation}\label{PS}
    P_{\Lambda\text{CDM}}(k)= \frac{C_{8}}{k^3}\ln^2{\left(\frac{k\tau_{\text{eq}}}{8}\right)},
\end{equation}
where $C_{8}$ is a constant fixed by amplitude of power spectrum. In our numerical calculation this formula works well for $k\gtrsim 1 (h/\text{Mpc})$. This corresponds to the scale of suppression from ULA with mass $m\gtrsim 10^{-24}$eV, which is the constrained scale of ULA mass from large scales observations if ULA is dominated. In this section we only consider this mass range. Then we can use transfer functions $T^2(k)$ we discussed in the last section to compute the variance in the presence of ULAs for $f\ll 1$.

\begin{figure}[tbp]
\centering
\includegraphics[scale=0.8]{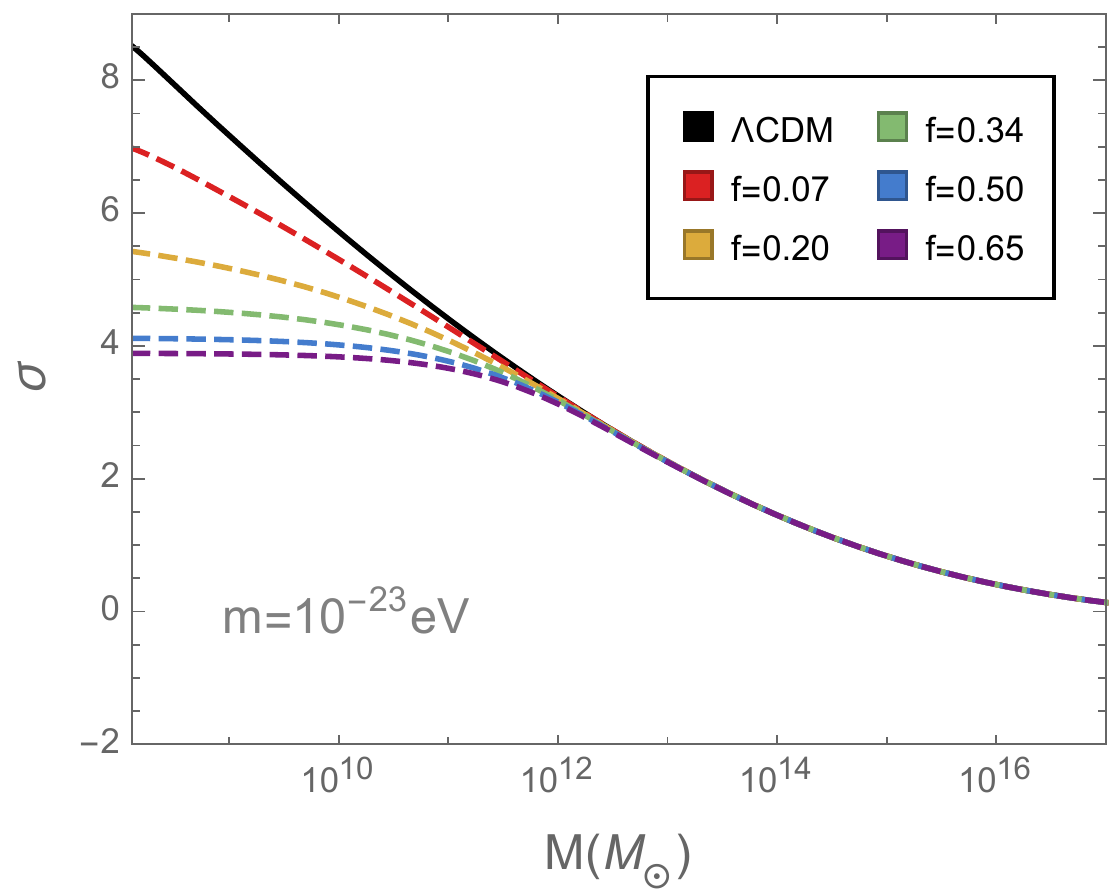}
\caption{\label{fig:Sig1}The variance for different fractions of ULA with the mass $m=10^{-23}$eV. The black curve is the $\Lambda$CDM case and the color curves are the ULA cases. We see the suppression starts at $10^{12}M_{\odot}$. For $f>0.5$, the suppression changes insignificantly.}
\end{figure}

We first consider the single-ULA with the mass $m=10^{-22}$eV. The power spectrum in the range $k<k_{m}$, corresponding to mass $M\gtrsim 10^{11}M_{\odot}$\footnote{In numerical calculations, we found the suppression starts actually at $\theta k_m$, where $\theta\sim0.6-0.7$. Hence the corresponding mass is close to $M\sim 10^{11}M_{\odot}$, rather than $10^{10}M_{\odot}$ estimated by (\ref{km}).}, is indistinguishable from $\Lambda$CDM. If we assume the flat primordial spectrum and the DM-dominated, then the leading order contribution to variance is \footnote{The numerical calculation result is not exactly consistent with this analytical formula without some corrections, such as the effects due to the BAO before recombination and the subleading contributions of the integral of window function. However, we here only discuss the mass function depending on ULAs approximately. }
\begin{equation}
    \sigma^2_{\Lambda\text{CDM}}(R)= \frac{C_{8}}{3}\ln^3{\left(\frac{R_{\text{eq}}}{8R}\right)},
\end{equation}
where $R_{\text{eq}}\equiv\pi\tau_{\text{eq}}\simeq 236.2(h^{-1}\text{Mpc})$. The variance of $\Lambda$CDM model is depicted as the black solid curve in Figure \ref{fig:Sig1}. The smaller regions have larger variance because the cluster starts earlier, which means the structure formation completed earlier. For the scale $k>k_{m}$ where the ULA starts to suppress the structure formation, the variance can be computed from (\ref{T1})
\begin{equation}\label{Sig1}
    \mathop{\sigma^2(R)}\limits_{R_{0}<R<R_{m}}= \Sigma(R; m,f)-\Sigma(R_{m}; m,f)+\sigma^2_{\Lambda\text{CDM}}(R_{m}),
\end{equation}
where we have denoted $R_{m}=\pi/k_{m}$, $R_{0}=\pi/k_{0}$ and 
\begin{equation}\label{sigma}
    \Sigma(R; m,f)=
    -C_8\left(\frac{R}{R_{m}}\right)^{24f/5}\sum_{n=0}^{2}\alpha_{n}f^{n-3}\ln^n{\left(\frac{R_{\text{eq}}}{8R}\right)},
\end{equation}
where 
\begin{equation}
\alpha_{0}=\frac{125}{6912},\ \ \ \  \alpha_{1}=\frac{25}{288},\ \ \ \   \alpha_{2}=\frac{5}{24}.
\end{equation}
We expect that the variance will be suppressed compared to $\Lambda$CDM for small mass haloes due to the suppression of small scale structure. In other words, ULA particles do not reside in structure for small scale because of its wave-like nature. To see this, we can compute
\begin{equation}\label{dsigma}
    \frac{d\sigma^2}{d\ln{M}}=-\left(\frac{R(M)}{R_{m}}\right)^{24f/5}\left[\frac{8 C_{8}}{5}\ln^2{\left(\frac{R_{\text{eq}}}{8R(M)}\right)}\right].
\end{equation}
For a fixed small mass, the variation of $\sigma^2$ is only exponentially sensitive to the relative relic energy of ULA $\ln{(|d\sigma^2/\ln{M}|)}\propto \Omega_{a}/\Omega_{m}$ for $f\ll 1$.  For $f\simeq0.5$, $d\sigma^2/d\ln{M}$ is small enough compared to $f=0$ hence the variance is flat enough for small structure formation, as one can see from dashed curves in Figure \ref{fig:Sig1}. For the fraction larger than $0.5$, the change of variance is not significant.

Starting from the analytical computation, we found the $\ln^2{M}$ term scales approximately as $M^{\delta}$ within the halo mass range we are interested in, where $\delta<0$. Therefore we can roughly fit $d\sigma^2/d\ln{M}$ for small halo mass as
\begin{equation}\label{dsigmaFit}
     \frac{d\sigma^2}{d\ln{M}}=\alpha\left(\frac{M}{4.78\times10^{10}M_{\odot}}\right)^{\gamma}\left[ m_{22}^{1.5}\left(\frac{M}{4.78\times10^{10}M_{\odot}}\right)\right]^{\lambda f},
\end{equation}
where $\alpha$, $\beta$, $\lambda$ and $\gamma$ are some parameters and can be fitted by numerical calculations. \footnote{For reference, these parameters are roughly fitted by $\alpha=-52.79$, $\lambda=1.13$ and $\gamma=-0.148$ and works for $10^{-24}\text{eV}\lesssim m \lesssim 5\times10^{-22}\text{eV}$ and fraction $0\lesssim f \lesssim 0.5$. These fitted parameters are not universal. Without the approximation, they actually depend on parameters of the models($n_s$, $\Omega_i$, $\tau_{\text{eq}}$...).} 
Although from the analytical results the $\ln{(|d\sigma^2/\ln{M}|)}$ scales as $\propto 8f/5$, which is only valid for small fraction, we find for $0\lesssim f \lesssim 0.5$ one can correct it to $\propto 8f/5-2f^2+2.2f^3$ to fit the numerical results. However, for rough estimations in $0\lesssim f \lesssim 0.5$ we found this can be replaced by linear $\propto \lambda f$. We show the results of (\ref{dsigmaFit}) as dashed curves in the Figure \ref{fig:Sig}.

\begin{figure}[tbp]
\centering
\includegraphics[scale=0.58]{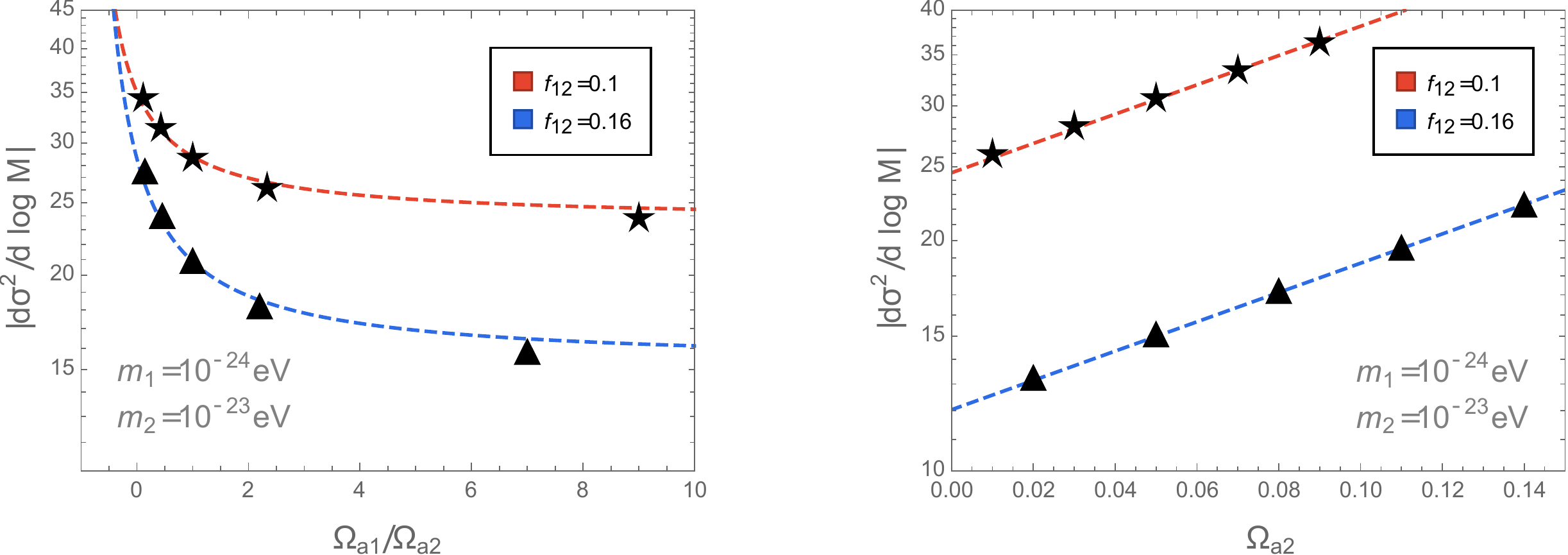}
\caption{\label{fig:Sigtwo}(Left) The derivative of variance for two ULAs v.s. $\omega_{r}=\Omega_{a1}/\Omega_{a2}$, at $7.73\times 10^{10}M_{\odot}$. (Right) The derivative of variance for two ULAs v.s. $\Omega_{a2}/\Omega_{m}$, at $2.75\times 10^9M_{\odot}$. The black stars and triangles are numerical results while the dashed curves are analytical results.}
\end{figure}

For two ULAs, if we consider $m_{r}\gtrsim0.017$, above the Jeans momentum $k_{m_{1}}$, the mass function is the same as the single-ULA case. Hence we can use (\ref{Sig1}) and (\ref{dsigma}) to compute the mass function for radius $R_{m_{2}}<R<R_{m_{1}}$. For $k>k_{m_{2}}$, the variance can be characterized by a single-ULA with effective mass $m_{e}$, which is exponentially sensitive to the relic density ratio of two ULAs $\omega_{r}$. Then the variance in the scale $R_{0}^{(1)}<R<R_{m_{2}}$ is given by
\begin{equation}
     \mathop{\sigma^2(R)}\limits_{R_{0}^{(1)}<R<R_{m_{2}}}= \Sigma(R; m_{e},f_{12})-\Sigma(R_{m_{2}}; m_{e},f_{12})+\sigma^2_{m_{1}}(R_{m_{2}}),
\end{equation}
where $\sigma^2_{m_{1}}(R)$ is the variance of the light ULA with mass $m_{1}$ given by (\ref{Sig1}). For a fixed  $f_{12}$, the mass function is exponentially sensitive to the relic density ratio $(1+\omega_{r})^{-1}$. Because in this range of halo mass $M$ with a size $R_{0}^{(1)}<R<R_{m_{2}}$ it is the same as an effective single-ULA, we can use (\ref{dsigmaFit}) to obtain the formula
\begin{equation}\label{CD}
    \frac{d\sigma^2}{d\ln{M}}\bigg{|}_{R_{0}^{(1)}<R<R_{m_{2}}}=C(M,m_{1},f_{12}) \left(\frac{1}{m_{r}}\right)^{\frac{1.5\lambda f_{12}}{1+\omega_{r}}}, \\
\end{equation}
where 
\begin{equation}
    C(M,m_{1},f_{12})\equiv \alpha\left(\frac{M}{4.76\times10^{10}M_{\odot}}\right)^{\gamma}\left[\left(\frac{m_{1}}{10^{-22}}\right)^{1.5}\left(\frac{M}{4.78\times10^{10}M_{\odot}}\right)\right]^{\lambda f_{12}}.
\end{equation}
The derivative of variance also scales as $(1+\omega_{r})^{-1}$ for mass $M$ between $M_{0}^{(1)}$ and $M_{m_{2}}$. In Figure \ref{fig:Sigtwo}(Left), we numerically calculate the $\ln{(|d\sigma^2/\ln{M}|)}$ with different $\omega_{r}$(black stars and black triangles), which agrees with (\ref{CD})(dashed curves).  We can see $|d\sigma^2/\ln{M}|$ decays more rapidly when $\omega_{r}\lesssim1$, while varies more slowly when the relative relic is larger than $1$. This is more significant for larger total ULA relic $f_{12}$, as the black triangles in Figure \ref{fig:Sigtwo}(Left) shows. This is because when $\omega_{r}\gtrsim 1$, the effective mass will be close to $m_{1}$. As we increase the amount of ULA-1, the variance also quickly approaches  to that of ULA-1. 

\begin{figure}[tbp]
\centering
\includegraphics[scale=0.8]{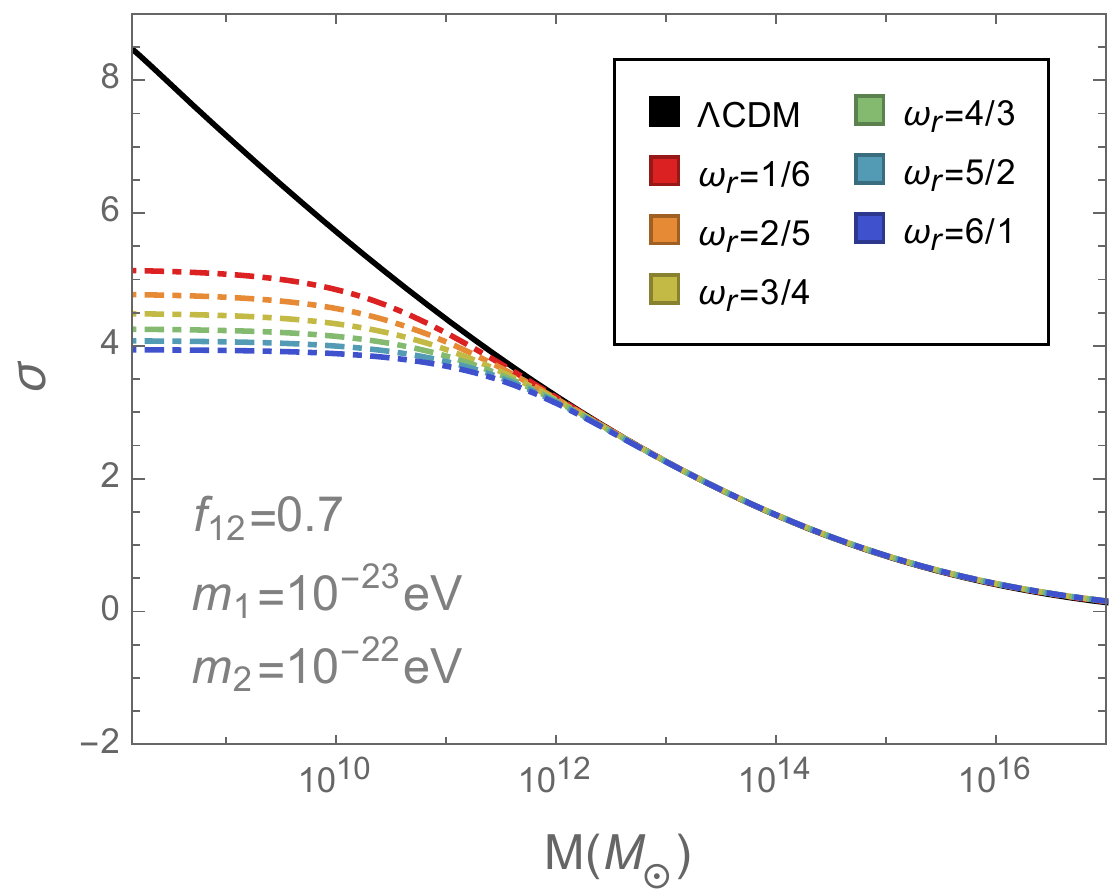}
\caption{\label{fig:Sig2}The variance for different relative relic $\omega_{r}$ and fixed $f_{12}=0.7$ of ULAs. The black curve is the $\Lambda$CDM case and the color curves are the ULAs cases. The variance are distinguishable for different $\omega_{r}$. And also the difference is more sensitive for $\omega_{r}\lesssim 1$.}
\end{figure}

Above $k_{0}^{(1)}$, the variance is given by
\begin{equation}\label{sigmal}
     \mathop{\sigma^2(R)}\limits_{R_{0}^{(2)}<R<R_{0}^{(1)}}= \Sigma(R; m_{n},f_{2})-\Sigma(R_{0}^{(1)}; m_{n},f_{2})+\sigma^2_{m_{e}}(R_{0}^{(1)}),
\end{equation}
where $\sigma^2_{m_{e}}(R)$ is the variance of the effective ULA with mass $m_{e}$ given by (\ref{Sig1}). This is because this term is just the boundary value at $R=R_{0}^{(1)}$, and soon above this scale the variance can be produced by the effective single-ULA. From (\ref{dsigma}), we can derive the derivatives of variance $\ln{(|d\sigma^2/\ln{M}|)\propto}f_{2}$ for a fixed $f_{12}$,
\begin{equation}\label{dsigmal}
    \frac{d\sigma^2}{d\ln{M}}\bigg{|}_{R_{0}^{(2)}<R<R_{0}^{(1)}}=D(M,f_{12})\left[\left(\frac{m_{2}}{10^{-22}}\right)^{1.5}\left(\frac{M}{8.71\times10^{7}M_{\odot}}\right)\right]^{\tilde{\lambda}f_{2}}, 
\end{equation}
where 
\begin{equation}
   D(M,f_{12})\equiv \tilde{\alpha}\left(0.017\right)^{1.5\tilde{\lambda}f_{12}} \left(\frac{M}{8.71\times10^{7}M_{\odot}}\right)^{\tilde{\gamma}}
\end{equation}
and $\tilde{\alpha}$, $\tilde{\lambda}$, and $\tilde{\gamma}$ are the parameters, which can be fitted by numerical calculations \footnote{For reference, we use $\tilde{\alpha}=128.88$, $\tilde{\lambda}=1.9$ and $\tilde{\gamma}=-0.148$.}. The number $0.017$ comes from  $\sqrt{\Omega_{r}/\Omega_{m}}$ and $f_{12}$ is the total fraction of ULAs.  As is shown in figure \ref{fig:Sigtwo}(Right), we can see for a fixed $f_{12}$, increasing $\Omega_{a2}$ will increase the steep of variance, i.e., $|d\sigma^2/d\ln{M}|$ increases. That means on these scales the variance has less suppression if we have larger $f_{2}$. This can be understood as following. From (\ref{lightT}), the transfer function on these scales can be written as $(\sqrt{0.017})^{24f_{1}/5}(k_{m_{2}}/k)^{24f_{2}/5}$, where the first factor is due to the suppression by ULA-1. We also know $\sqrt{0.017}<k_{m_{2}}/k<1$. Therefore when we increase $f_{1}$, at the same time decrease $f_{2}$ and keep the $f_{12}$ unchanged, the mitigation of suppression of ULA-2 $(k_{m_{2}}/k)^{24f_{2}/5}$ can not offset the aggravation of suppression of ULA-1 $(\sqrt{0.017})^{24f_{1}/5}$, which results in the flatter $\sigma^2$ for smaller $f_{2}$. We show the results of variance with different $\omega_r$ in Figure (\ref{fig:Sig2}). It's shown that the variance is distinguishable even when the ULAs is the dominated dark matter.

For the case $m_{r}\lesssim0.017$, the effects of two ULAs are almost independent. The suppression of ULA-1 is the same as single-ULA case. Then one can use (\ref{sigmal}) to compute the variance of ULA-2 by replacing $R_{0}^{(1)}\to R_{m_{2}}$ and $\sigma^2_{m_{e}}\to \sigma^2_{m_{1}}$. The derivative of variance is the same as (\ref{dsigmal})  between $R_{0}^{(2)}<R<R_{m_{2}}$.

\subsection{Halo mass function}
\label{sec:mf}

\begin{figure}[tbp]
\centering
\includegraphics[scale=0.55]{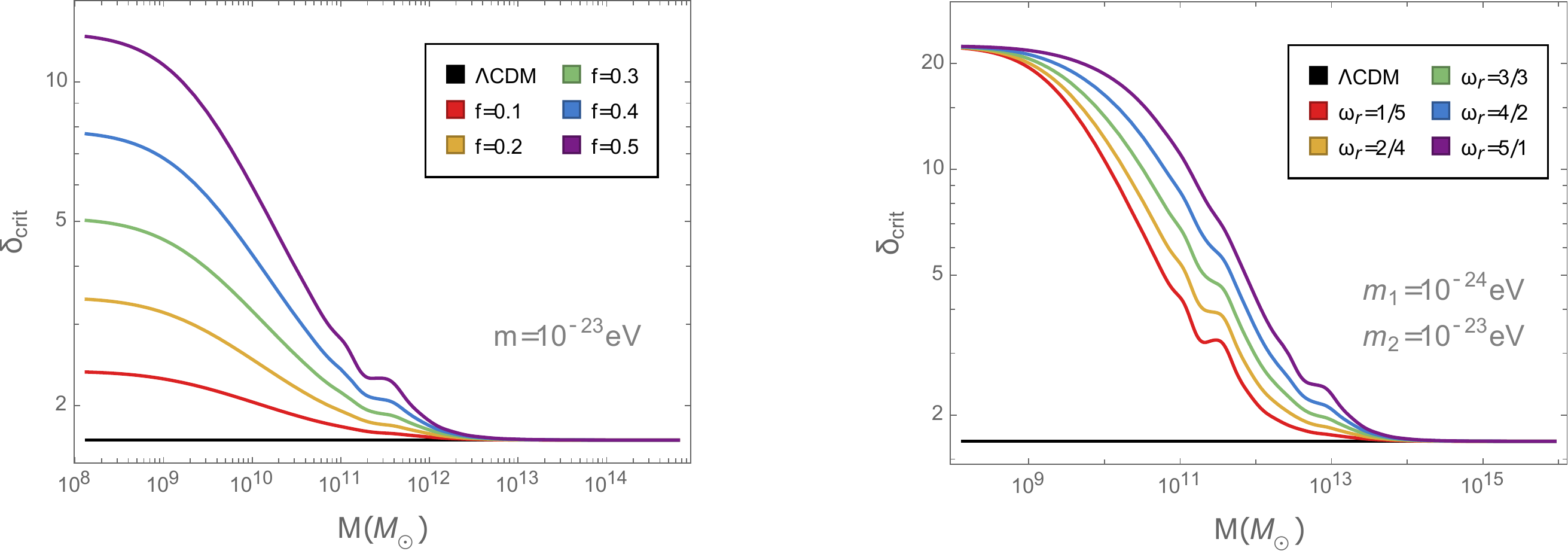}
\caption{\label{fig:CDt}(Left)The critical density $\delta_{\text{crit}}$ in first-crossing function for different fraction $f$ of single-ULA with mass $m=10^{-23}$eV. (Right)The critical density $\delta_{\text{crit}}$ in first-crossing function for different relative fraction $\omega_{r}$ of two ULAs with mass $m_{1}=10^{-24}$eV, $m_{2}=10^{-23}$eV and $f_{12}=0.6$. The black curves are the one in $\Lambda$CDM $\delta_{\text{crit}}^{\Lambda\text{CDM}}$=1.686.}
\end{figure}

To calculate the mass function, we need to know the cut-off mass of the collapsed objects.  The collapse is significantly suppressed if WDM or ULAs are  dominant component of dark matter. If we have a mass scale of free-streaming of dark matter $M_{\text{fs}}$, the mass function will drop down rapidly for $M<M_{\text{fs}}$. For $\Lambda$CDM model, the critical density is a constant $\delta_{\text{crit}}^{\Lambda\text{CDM}}\simeq 1.686$ for all size or mass of collapsed objects. If we apply this critical density to the WDM or ULAs, the mass function $dn/d\ln{M}$ scales as $R^{-1}$ for small mass objects, and consequently diverges. A sharp-$k$ window function is used in the WDM and ULAs models, which can provide the results in good agreement with simulations for the low-mass haloes \cite{Schneider:2013ria,Schneider:2014rda,Kulkarni:2020pnb}. However, in the absence of simulations, it is not clear which cut-off scale should be taken in the sharp-$k$ filter, especially for the multi-ULA case. Since we are considering a mixing ULAs and CDM universe, the structure suppression is not so significant on small scales.  The mixing of different ULAs gives rise to the suppression of the power spectrum on different scales. Therefore here we prefer to use the top-hat filter and adopt a mass dependent critical density $\delta_{\text{crit}}(M)$ in the first-crossing function \cite{Marsh:2013ywa,Bozek:2014uqa,Du:2016zcv}\footnote{Although in \cite{Du:2016zcv} the authors claimed that the first-crossing function of $\Lambda$CDM is not applicable to ULAs because we use the different critical density. We here does not do the simulations. We only study how modification of mass function depends on the multiple ULAs.}.

The scale-dependent critical density is considered due to the scale-dependent growth factor $D(k,a)$ of dark matter in the matter dominant era, as one can see in Figure \ref{fig:overdensity}. The growth factor is suppressed on small scales because of the free-streaming of ULAs. One can define a ratio of the growth factors as \cite{Marsh:2013ywa}
\begin{equation}
    \mathcal{G}(k,a)\equiv \frac{D_{\Lambda\text{CDM}}(a)}{D_{\text{ULAs}}(k,a)}.
\end{equation}
To collapse at the same redshift $z_{c}$ as the $\Lambda$CDM universe, the critical density of ULAs should be larger than $\Lambda$CDM so that the overdensity can reach $1.686$ at $z_{c}$, i.e., 
\begin{equation}
    \delta_{\text{crit}}=\mathcal{G}(k,a)\delta_{\text{crit}}^{\Lambda\text{CDM}}.
\end{equation}
After normalizing such that $D(k=k_{\text{pivot}},a=1)=1$, the suppression factor $\mathcal{G}(k,a=1)$ at present time can be deduced as
\begin{equation}
    \mathcal{G}(k)=\frac{\delta_{\Lambda\text{CDM}}(k,1)\delta_{\Lambda\text{CDM}}(k_{\text{pivot}},a_{h})}{\delta_{\Lambda\text{CDM}}(k,a_{h})\delta_{\Lambda\text{CDM}}(k_{\text{pivot}},1)} \frac{\delta_{\text{ULAs}}(k,a_{h})\delta_{\text{ULAs}}(k_{\text{pivot}},1)}{\delta_{\text{ULAs}}(k,1)\delta_{\text{ULAs}}(k_{\text{pivot}},a_{h})},
\end{equation}
where $a_{h}$ should be chosen in $\Lambda$CDM universe as the time near recombination, where the BAO does not affect the dark matter overdensity so that the transfer function has frozen in. In our code we find $a_{h}\simeq 0.0035$ is suitable. We show the critical density of single-ULA for different fractions in Figure \ref{fig:CDt}(Left). For larger $f$, the suppression of growth factor on small scales is stronger (larger $\mathcal{G}$) hence we have a larger critical density. For two ULAs, we also show $\delta_{\text{crit}}$ in Figure \ref{fig:CDt}(Right) for different $\omega_{r}$ with a fixed $f_{12}$. Because for heavy ULA, the suppression starts on smaller scales. And the critical density starts to distinguish from $\delta_{\text{crit}}^{\Lambda\text{CDM}}$ on smaller mass of collapsed objects. Therefore when increasing $\omega_{r}$ the effective mass of ULA is closer to one of the light ULA so the critical density is different from $1.686$ on larger $M$. This means the cut-off mass is larger for larger $\omega_{r}$.

\begin{figure}[tbp]
\centering
\includegraphics[scale=0.75]{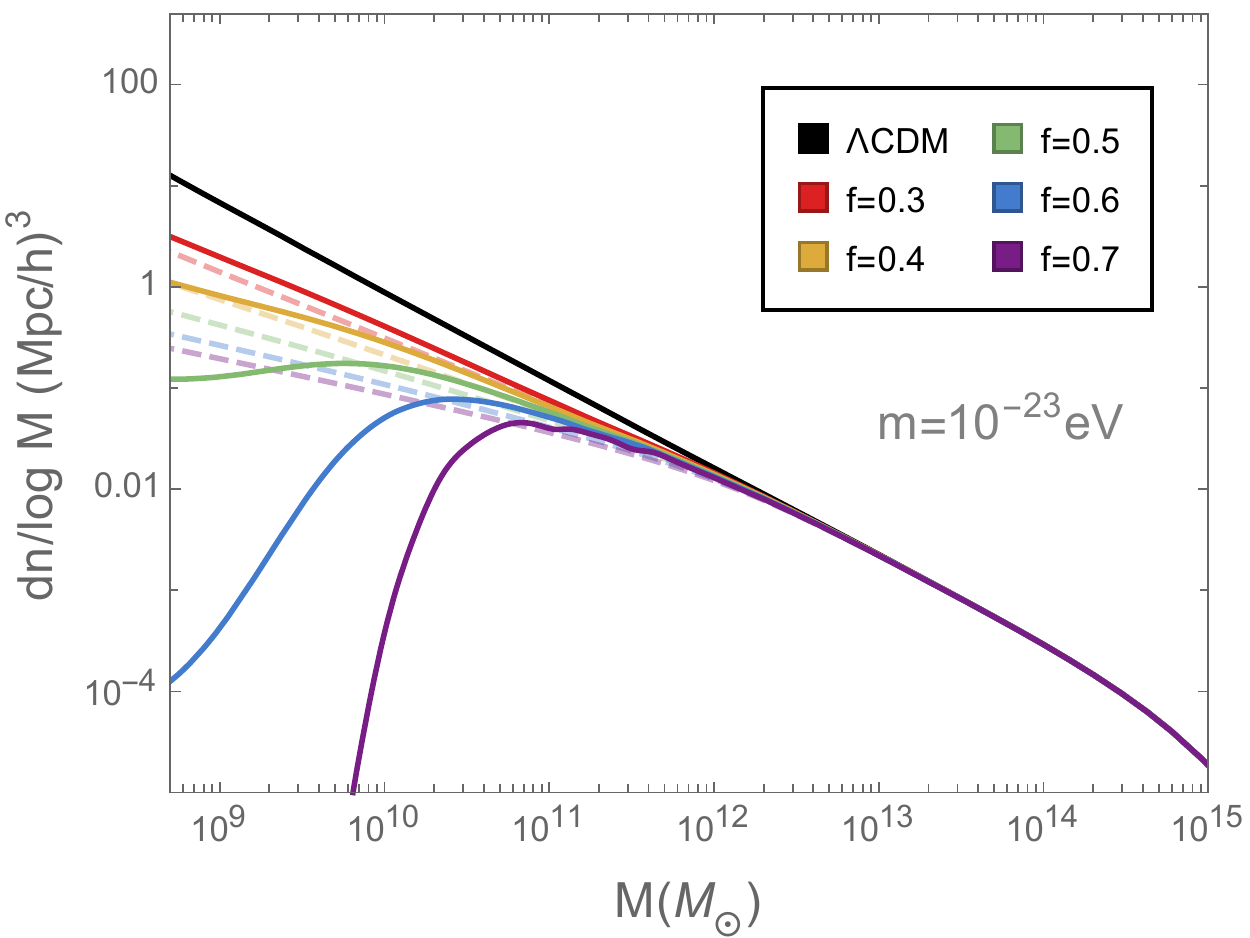}
\caption{\label{fig:MFone}We plotted the mass functions of single-ULA with mass $10^{-23}$eV. The solid curves represent those with the scale-dependent critical overdensity $\delta_{\text{crit}}(M)$, while the dashed curves represent those with the $\Lambda$CDM critical overdensity $\delta_{\text{crit}}^{\Lambda\text{CDM}}=1.686$.}
\end{figure}

\begin{figure}[tbp]
\centering
\includegraphics[scale=0.75]{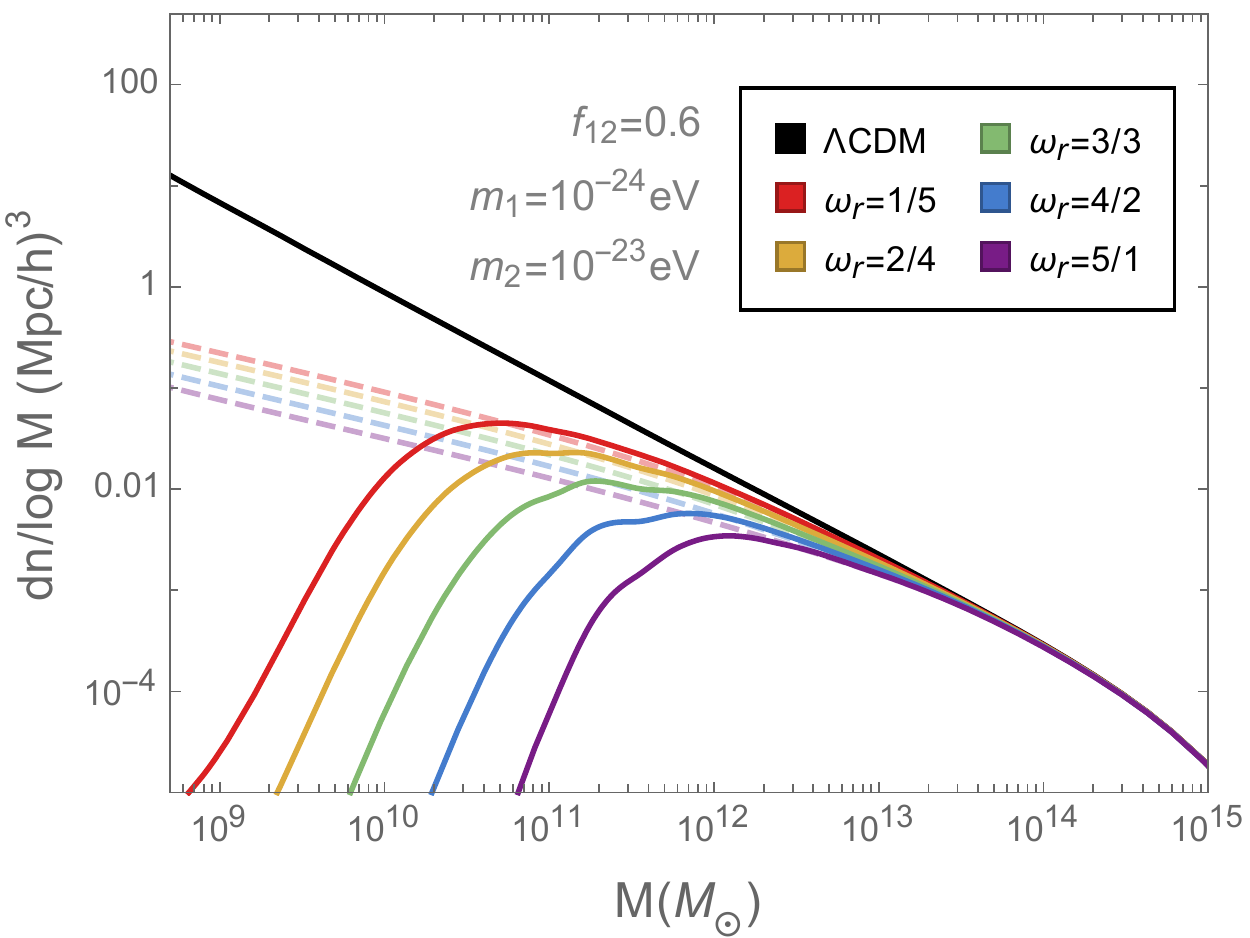}
\caption{\label{fig:MFtwo}The mass function of two ULAs with mass $m_{1}=10^{-24}$eV, $m_{2}=10^{-23}$eV and total relic $\Omega_{a}=0.6\Omega_{m}$. The solid curves are the mass function  with the scale-dependent critical overdensity $\delta_{\text{crit}}(M)$, while the dashed curves are that with the $\Lambda$CDM critical overdensity $\delta_{\text{crit}}^{\Lambda\text{CDM}}=1.686$.}
\end{figure}

We show the mass function of single-ULA in Figure \ref{fig:MFone}. To understand the dependence of the mass function on ULA mass and fraction, we need to know 
\begin{equation}
    \mathcal{N}(M)\equiv -\frac{1}{2}\frac{\rho_{m}}{M\sigma^2}\frac{d\sigma^2}{d\ln{M}}.
\end{equation}
We have learned the derivative $d\sigma^2/d\ln{M}$ in the last subsection. It scales as $M^{\gamma+\lambda f}E^{\lambda f}$ in the haloes mass range we are interested in, where $E$ is constant for fixed $M$. After integrating we obtain the $\Sigma$, which scales as $F(f)M^{\gamma+\lambda f}E^{\lambda f}$. From (\ref{sigma}) we can know $F(f)$ are just polynomials of $f$ and logarithmic dependence of $M$. After ignoring the dependence of $f$ of these polynomials, we can find the function $\mathcal{N}$ for $f<0.5$ roughly scales as 
\begin{equation}
    \mathcal{N}=\epsilon \left(\frac{M}{7.27\times10^{10}M_{\odot}}\right)^{\xi}\left[\left(\frac{m}{10^{-22}}\right)^{1.5}\frac{M}{7.27\times10^{10}M_{\odot}}\right]^{\zeta f},
\end{equation}
where $\epsilon$, $\xi$ and $\zeta$ are some positive parameters fixed by numerical calculations\footnote{For reference, we use $\epsilon=0.69$, $\xi=-0.93$ and $\zeta=0.72$.}.
For two-ULA case, we can just replace the $m$ as the effective single-ULA, similarly to the variance we discussed above. That is, on scales where two ULAs suppress the structure formation, the function $\mathcal{N}$ scales as $(1/m_{r})^{(\zeta f)/(1+\omega_{r})}$. Therefore we expect $\mathcal{N}$ is more sensitive for $\omega_{r}\lesssim1$ when the halo mass is greater than the cut-off mass.

\begin{figure}[tbp]
\centering
\includegraphics[scale=0.5]{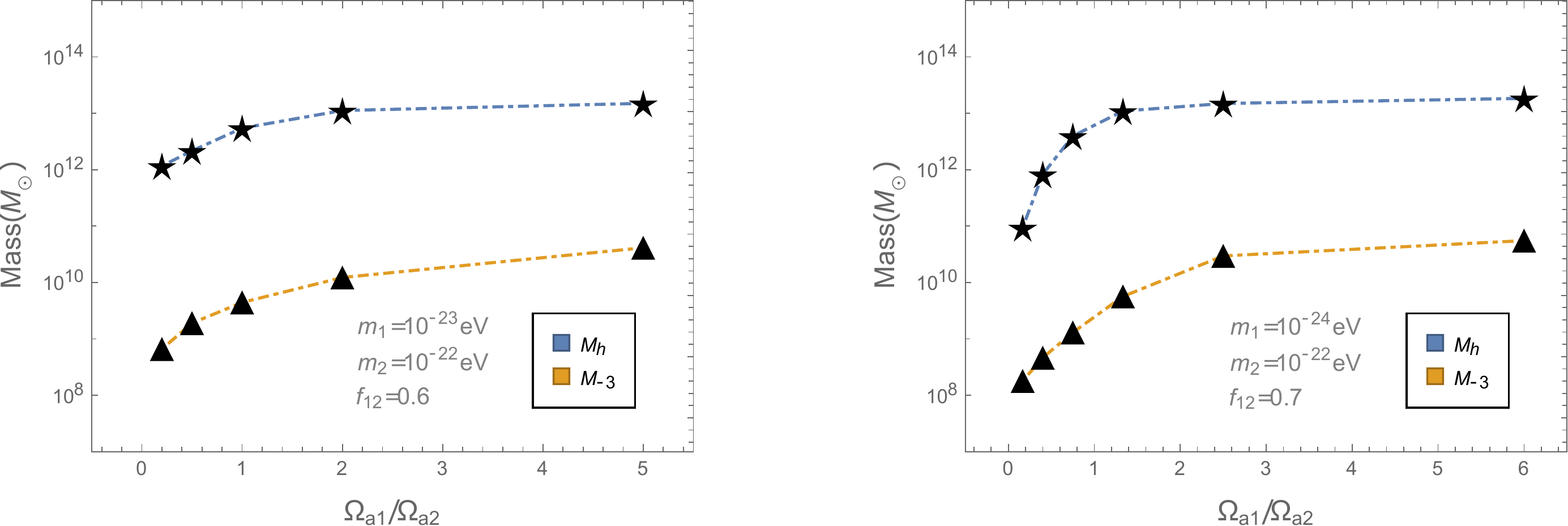}
\caption{\label{fig:Mcut}The half-mode mass $M_{h}$ and cut-off mass $M_{-3}$ of collapsed objects for two-ULA universe with $m_{r}=0.1$(Left) and $m_{r}=0.01$(Right). $M_{-3}$ is three orders of magnitude smaller than $M_{h}$.}
\end{figure}

For the single-ULA, the cut-off mass is at a mass intermediate between the Jeans mass and the half-mode mass \cite{Marsh:2013ywa}. We see in Figure \ref{fig:MFone} for $f>0.5$, the mass function rapidly falls on small scales. For two ULAs, the situations are more complex because we have another two parameters $m_{r}$ and $\omega_{r}$ in the model. Even if we fix masses of two ULAs, the half-mode scale is moving with different relative relic $\omega_{r}$. Fortunately,  we have an argument that the half-mode scale of two-ULA can be estimated by one of the effective single-ULA. Therefore we expect that the cut-off scale is at scale below $M_{h}$. And the half-mode mass can be estimated by effective ULA mass as $M_{h}\simeq m_{e}^{-1.35}$ \cite{Marsh:2013ywa}. To see the cut-off mass scale for the two ULAs, we defined the mass scale $M_{-3}$ as the scale that the number of haloes has dropped to a thousandth of that at half-mode mass, i.e.,
\begin{equation}
    \frac{dn(M_{-3})}{d\ln{M}}=10^{-3} \frac{dn(M_{h})}{d\ln{M}}.
\end{equation}
Below the mass $M_{-3}$, the number of haloes is small enough compared to the number of haoes with half-mode mass. We show these mass scales in Figure \ref{fig:Mcut}. We choose $f_{12}>0.5$ so that the cut-off is significant. For $m_{r}=0.1$ and $f_{12}=0.6$, we see the half-mode mass is about $M_{h}\gtrsim 10^{12}M_{\odot}$. In Figure \ref{fig:MFtwo}, this mass scale is still far from the cut-off.
Hence, as we the mass decreases, the number of haloes increases. As we have mentioned before, the half-mode mass is more sensitive when $\omega_{r}\lesssim 1$. For this case, scale $M_{-3}$ is about $M_{-3}>10^{9}M_{\odot}$, three orders of magnitude smaller than $M_{h}$. Also, this mass is sensitive in the range $\omega_{r}\lesssim 1$, similarly to the half-mode mass. For $m_{r}=0.01$ and $f_{12}=0.7$, the half-mode mass varies larger order of magnitude than the $m_{r}=0.1$ case, as we mentioned before. However, the mass scale $M_{-3}$ still varies one order of magnitude between $1<\omega_{r}<3$, while the half-mode mass varies insignificantly in this range. This may be due to the larger mass hierarchy so that the suppression of two ULAs are almost independent. When $\omega_{r}>1$, at some scales after half-mode scale the suppression has frozen in because the effect of ULA-2 has not been relevant. Hence the $M_{-3}$ is at smaller scales than we expected. The half-mode scale is probably not a good characteristic scale for this case. 

\begin{figure}[tbp]
\centering
\includegraphics[scale=0.49]{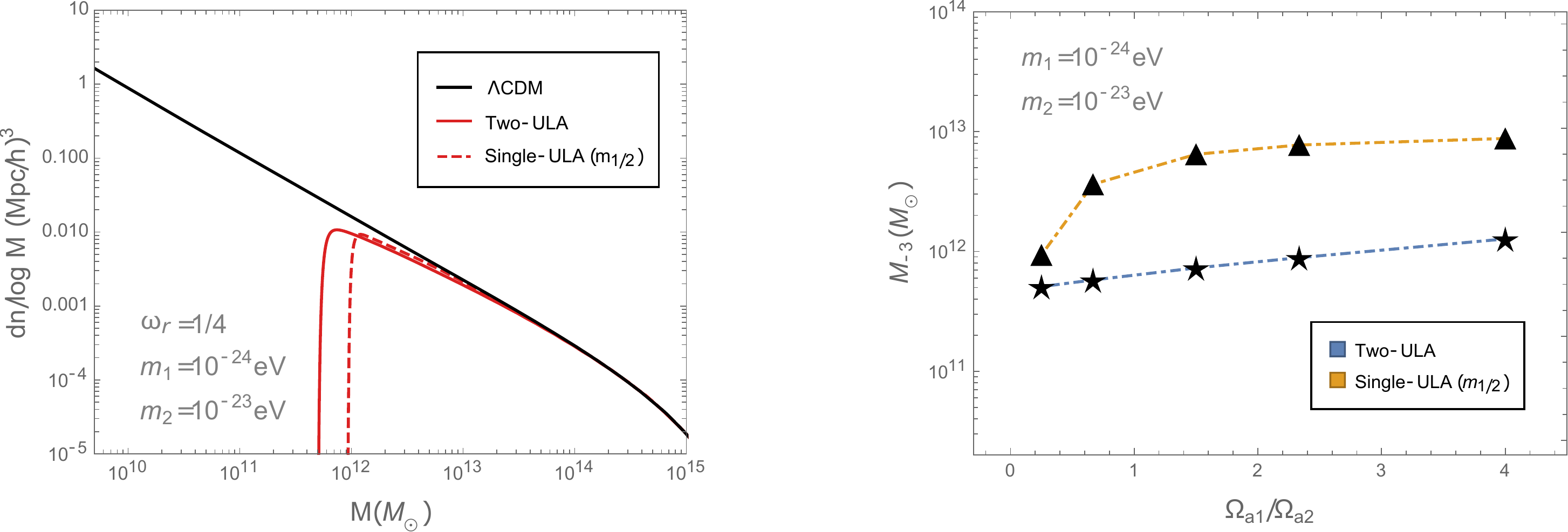}
\caption{\label{fig:Mc}(Left) The mass functions when all dark matter are ULAs. The red solid curve is the mass function of two-ULA with $\omega_{r}=1/4$, while the dashed curves is that of single-ULA, which share the same half-mode scale with the two-ULA case. (Right) The cut-off mass $M_{-3}$ for two-ULA case and single-ULA case sharing same half-mode scale.}
\end{figure}

We have seen that for $m_{r}>0.017$, the effective single-ULA catches some important features of the two ULAs. Because the overlap suppression interval of two ULAs can be reproduced by the effective single ULA for small $f_{12}$. The question is if we can distinguish the mass function of two ULAs  from that of single-ULA. We have shown that the linear power spectrum of these two cases are quite different. We here calculate the mass functions for two-ULA and the single-ULA with mass (\ref{m1/2}), which share the same half-mode mass. As Figure \ref{fig:Mc} shows, we consider $f_{12}=1$, where the cut-off mass is visible. We also use $M_{-3}$ as the cut-off mass. The mass functions of these two case are quite different. For example for $\omega_{r}=1/4$, before the cut-off mass, the number of haloes for two-ULA case is $10\%$ smaller than the single-ULA case, due to the suppression of the light ULA. And the cut-off mass of two ULAs $\sim 5\times 10^{11}M_{\odot}$ is smaller than the single-ULA $\sim 10^{12}M_{\odot}$, due to the suppression of the heavy ULA.

\section{Conclusion}
In this paper, we have systematically studied the implications of multi-ULA on the formation of the large scale structure. We specifically investigated the two-ULA case for understanding, where their masses and initial conditions can be quite different. We denote these two ratio parameters as $m_{r}\equiv m_{1}/m_{2}$ and $\omega_{r}\equiv \Omega_{a1}/\Omega_{a2}$ (in this paper we always set $m_{1}<m_{2}$). We also developed a numerical code to calculate the power spectrum with the introduction of the new fluid approximation method \cite{Passaglia:2022bcr}, and used them to check the analytical results.

Since the suppression scale of ULA dark matter in the power spectrum depends on the mass of ULAs, we discuss two separate cases: $m_{r}\lesssim 0.017$ and $m_{r}\gtrsim 0.017$. For the former case, the suppression of two ULAs are almost independent hence we can treat them separately. For the latter case, there is an overlap region where both two ULAs suppress the structure formation. We found on these scales the suppression can be reproduced by an effective single ULA with the effective mass $m_{e}$ and the effective relic density parameter $\Omega_{e}=\Omega_{1}+\Omega_{2}$. Specifically, for small $\Omega_{e}$ the results can be computed analytically and this effective mass is given by $\sim m_{r}^{-1/(1+g(\omega))}$, where $g(\omega_{r})\simeq \omega_{r}$ for small $\Omega_{e}$. 
We have also obtained a general formula for the effective mass in multi-ULA cases. 
From the relation, we learned that the power spectrum is more sensitive when $\omega_{r}\lesssim 1$. We naively compare the results with constraints on WDM when all dark matter is composed of free-streaming dark matter. When $\omega_{r}\gtrsim 1$ the power spectrum rapidly reduce to that of the light ULA. This provides a strong constraint on the relic of the light ULA. We found if the light ULA has the mass $10^{-22}$eV, the amount of heavy ULAs should be much more than that of the light one if the constrain of Lyman-$\alpha$ forest is imposed. The relics of multiple ULAs also determine the structure of galaxies when they form condensates. If self-interaction is relevant, there exist stable solutions in the parameter space for this situation \cite{Eby:2020eas}.

Moreover, we studied the haloes mass function of the two-ULA universe. We approximately compute the variance and also the mass function as a function of two parameters $m_{r}$ and $\omega_{r}$. We extended the mass-dependent critical density of collapsed objects to the two-ULA case and found the mass function is also more sensitive for $\omega_{r}\lesssim 1$ for mixing models, due to the effective single-ULA effect. Since the linear power spectrum of two-ULA case is  different from that of the single-ULA which share the same half-mode scale,  the mass functions of them are  distinguishable. The mass function of two ULAs has less haloes and smaller cut-off mass.

Although we compared the results with the WDM constraints, we have mentioned that this comparison is not accurate because the mass function is sensitive to the shape of the linear power spectrum. For multi-ULA case, these shapes are various and in general it is not easy to find a characteristic scale. In principle, we can do non-linear simulations for catching these feature of mass functions of multi-ULA universe. However, a self-consistent simulation is still lacking (one can find a summary of current simulations for ULA in \cite{Zhang:2018ghp}). To address the mass functions, the low-cost hydrodynamic simulations have been developed by different groups. The most important effect is the quantum pressure of the ULAs. In \cite{Irsic:2017yje} this quantum effect is ignored because the scales of data they used is smaller than the ULA Jeans scale. However, it has been pointed out that the quantum pressure may be important for the structure formation \cite{Zhang:2017chj}. It is risky to ignore this quantum pressure in the equations. The simulation is still understudied hence in this paper we provided some insights of structure formation of multi-ULA before fully understanding the hydrodynamic simulations. Also, the simulation of mixed CDM is also performed recently \cite{Vogt:2022bwy,Schwabe:2020eac}. To construct the density profile of dark matter haloes, the mixing with small fraction of CDM is also important \cite{Marsh:2013ywa}. We leave these issues for future work.

\acknowledgments

We would like to thank Pak Hang Chris Lau and Samuel Charles Passaglia for helpful discussions. J.\ S. was in part supported by JSPS KAKENHI Grant Numbers JP17H02894, JP17K18778, JP20H01902, JP22H01220. C-B.\,C. was supported by Japanese Government (MEXT) Scholarship and China Scholarship Council (CSC).

\appendix
\section{Accurate matching of ULAs}
\label{IC}
In this appendix, we provide the initial conditions from the matching conditions between scalar field and effective fluid of ULAs at a switching time, which are proposed in \cite{Passaglia:2022bcr}. We here show how to obtain these initial conditions in our code.

The solutions of ULA scalar field can be written as
\begin{equation}
    \phi(t)=\varphi_{c}(t)\cos{\left[m(t-t_{*})\right]}+\varphi_{s}(t)\sin{\left[m(t-t_{*})\right]}.
\end{equation}
where $t_{*}$ is the switching time between scalar field and effective fluid. It is convenient to calculate numerically with defining variable $y=a(\tau)$ and in this paper we choose the switching time at $my_{*}/\mathcal{H}_{*}=12$. Denoting dot as derivative of $mt$. One can obtain the equations of $\varphi_{c}$ and $\varphi_{s}$ from K-G equation
\begin{align}
    &\ddot{\varphi}_{c}+2\dot{\varphi}_{s}+\frac{3\mathcal{H}}{my}\left(\varphi_{s}+\dot{\varphi}_{c}\right)=0,\nonumber\\
    &\ddot{\varphi}_{s}-2\dot{\varphi}_{c}+\frac{3\mathcal{H}}{my}\left(-\varphi_{c}+\dot{\varphi}_{s}\right)=0,
\end{align}
After the switching time, where ULAs have started to oscillate, the scalar fields evolve as dust, which means $\varphi_{c,s}\sim a^{-3/2}$. Then one can impose the matching conditions at the switching time
\begin{equation}\label{varphi}
    D_{*}\equiv\frac{\ddot{\varphi}_{c,s}}{\dot{\varphi}_{c,s}}\bigg{|}_{*}=-\frac{\mathcal{H}_{*}}{2my_{*}}\left[3-\frac{2y_{*}}{\mathcal{H}_{*}}\left(\frac{d\mathcal{H}_{*}}{dy}-\frac{\mathcal{H}_{*}}{y_{*}}\right)\right]
\end{equation}
Using these conditions we can reduce the equations (\ref{varphi}) to two first-order equations. If we consider the equations at the switching time, then we have two equations of $\varphi_{c*}$, $\varphi_{s*}$, $\dot{\varphi}_{c*}$ and $\dot{\varphi}_{s*}$. Also, at the switching time, we have $\phi(t_{*})\equiv\phi_{*}$ and $d\phi(t_{*})/dt\equiv\dot{\phi}_{*}$, which imply another two matching conditions
\begin{equation}
    \varphi_{c*}=\phi_{*}, \ \ \ \ \ \dot{\varphi}_{c*}+\varphi_{s*}=\frac{{\phi}_{1*}}{m}.
\end{equation}
Combining the above four equations, one can solve the quantities at the switching time 
\begin{align}
    &\varphi_{c*}=\phi_{*},\nonumber\\
    &\varphi_{s*}=-\frac{-6\mathcal{H}_{*}\phi_{*}/y_{*}-\left(4+D_{*}^2+6D_{*}\mathcal{H}_{*}/my_{*}+9\mathcal{H}_{*}^2/m^2y_{*}^2\right)\dot{\phi}_{*}}{m\left(4+D_{*}^2+3D_{*}\mathcal{H}_{*}/my_{*}\right)},\nonumber\\
    &\dot{\varphi}_{c*}=-\frac{\left(3\mathcal{H}_{*}/my_{*}\right)\left(2m\phi_{*}+D_{*}\dot{\phi}_{*}+3(\mathcal{H}_{*}/my_{*})\dot{\phi}_{*}\right)}{m\left(4+D_{*}^2+3D_{*}\mathcal{H}_{*}/my_{*}\right)},\nonumber\\
    &\dot{\varphi}_{s*}=\frac{\left(3\mathcal{H}_{*}/my_{*}\right)\left(D_{*}m\phi_{*}-2\dot{\phi}_{*}\right)}{m\left(4+D_{*}^2+3D_{*}\mathcal{H}_{*}/my_{*}\right)},
\end{align}
where all quantities can be represented by $\phi_{*}$, $\dot{\phi}_{*}$ and $\mathcal{H}_{*}$. The conformal Hubble variable at the switching time can be computed by Friedmann equation
\begin{equation}
    \mathcal{H}_{*}^2=\mathcal{H}_{0}^2y_{*}^2\left[\Omega_{r}/y_{*}^4+\left(\Omega_{m}-\Omega_{a}\right)/y_{*}^3+\Omega_{\Lambda}+\left(\frac{1}{2}\phi_{1*}^2+\frac{m^2}{2}\phi_{*}^2\right)/\rho_{cr}\right].
\end{equation}
Therefore the quantities $\varphi_{c*}$, $\varphi_{s*}$, $\dot{\varphi}_{c*}$ and $\dot{\varphi}_{s*}$ are only related to $\phi_{*}$ and $\dot{\phi}_{*}$. 

\begin{figure}[tbp]
\centering
\includegraphics[scale=0.6]{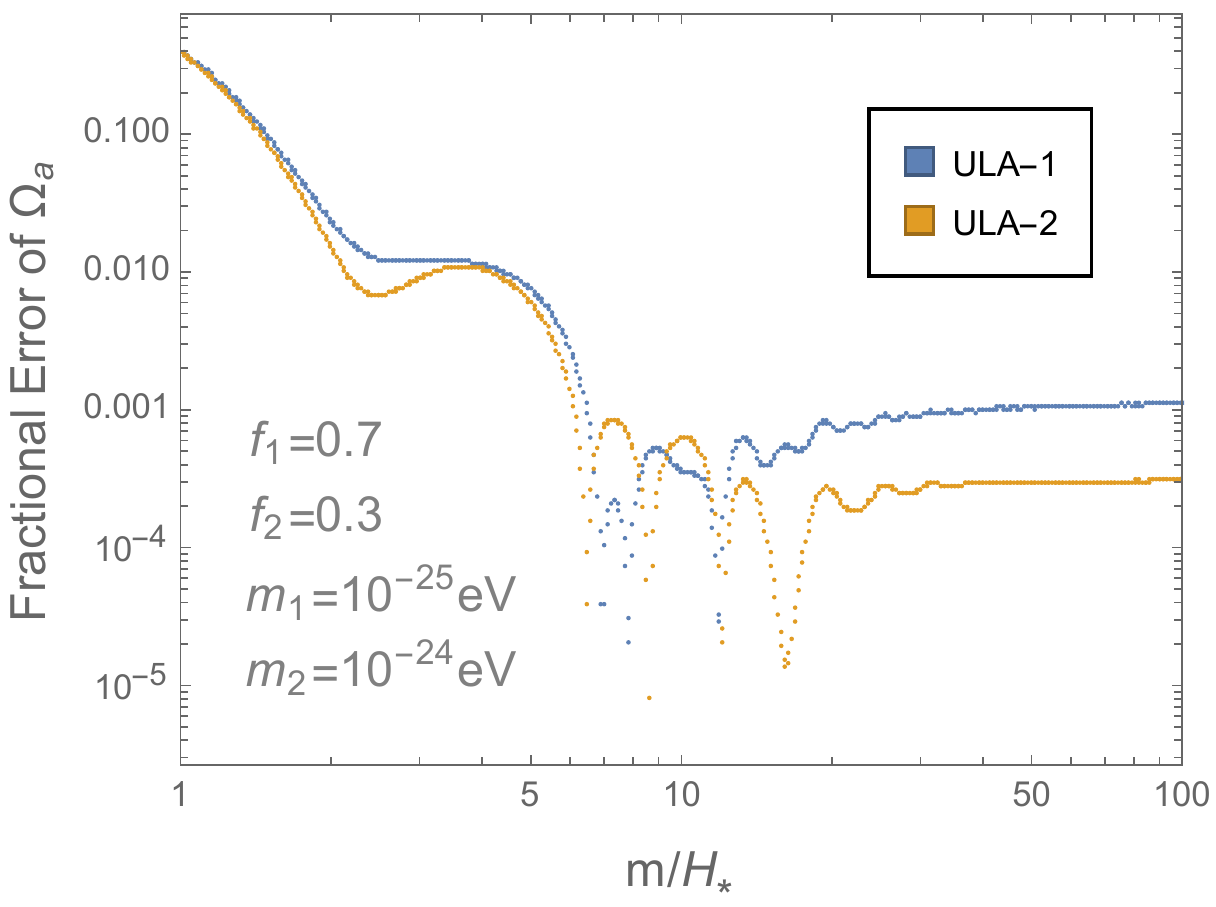}
\caption{\label{fig:error}The fractional errors of relic density of two ULAs of the shooting method in the code. We show an example where $m_{1}=10^{-25}$eV, $m_{2}=10^{-24}$eV and $f_{1}=0.7$, $f_{2}=0.3$. In this paper, we used $m/H_{*}\sim 10$, which reduces $\sim0.01\%$ fractional error.}
\end{figure}

On the other hand, after averaging the oscillation of ULA, the energy density of effective fluid of ULA at the switching time can be constructed by
\begin{equation}
    \rho_{*}^{\text{eff}}=\frac{1}{2}m^2\left(\varphi_{c*}^2+\varphi_{s*}^2+\frac{\dot{\varphi}_{c*}}{2}+\frac{\dot{\varphi}_{s*}}{2}-  \varphi_{c*}\dot{\varphi}_{s*}+\varphi_{s*}\dot{\varphi}_{c*}\right),
\end{equation}
which can be represented by $\phi_{*}$ and $\dot{\phi}_{*}$. If we assume the relic of ULAs today is $\Omega_{a}$, one can use the evolution of effective ULA fluid by using (\ref{axioneffeom}) and (\ref{axioneffeom1}) with effective equation of state and sound speed (\ref{weffceff}) from today to the switching time and obtain the density $\rho_{*}^{\text{eff}}$, which is only related to $\phi_{*}$ and $\dot{\phi}_{*}$. Then we can assume that the initial conditions of ULAs are $\phi_{i}=C$ and $\phi_{i}'=0$ and use the K-G equation to obtain the correct $C$ which shoots the desired density $\rho_{*}^{\text{eff}}(\phi_{*},\dot{\phi}_{*})$ at the switching time. The accuracy of this method depends on the switching time and we choose $m/H_{*}=12$ which results in fractional error of $\Omega_{a}$ into $10^{-4}\sim 10^{-3}$, see Figure \ref{fig:error}.


\end{document}